\documentclass[10pt]{article}
\usepackage{amsmath}
\usepackage{graphicx}
\usepackage{color}
\usepackage{orcidlink}
\usepackage{hyperref}
\usepackage{multirow}
\usepackage{rotating}
\usepackage{multicol}
\hypersetup{colorlinks=true, linkcolor=blue, citecolor=blue, urlcolor=blue}
\usepackage{caption}
\usepackage{subcaption}
\usepackage{setspace}
\RequirePackage[numbers,sort&compress]{natbib}
\begin{document}
\baselineskip=18pt

\begin{center}
\LARGE{{{Phase Transitions, Geodesic Structure, and Thermodynamic Properties Measurement of Einstein-Maxwell-Power Yang-Mills Black Hole Models}}}
\par\end{center}

\vspace{0.3cm}

\begin{center}
{\bf Abdelmalek Bouzenada\orcidlink{0000-0002-3363-980X}}\footnote{\bf abdelmalekbouzenada@gmail.com}\\ 
   {\tt Laboratory of Theoretical and Applied Physics, Echahid Cheikh Larbi Tebessi University 12001, Algeria}\\
    {\tt Research Center of Astrophysics and Cosmology, Khazar University, Baku, AZ1096, 41 Mehseti Street, Azerbaijan}\\
\vspace{0.1cm}
{\bf Allan. R. P. Moreira\orcidlink{0000-0002-6535-493X}}\footnote{\bf allan.moreira@fisica.ufc.br }\\ 
{\tt Secretaria da Educa\c{c}\ {a}o do Cear\'{a} (SEDUC), Coordenadoria Regional de Desenvolvimento da Educa\c{c}\ {a}o (CREDE 9),  Horizonte, Cear\'{a}, 62880-384, Brazil}\\
\vspace{0.1cm}
{\bf Shi-Hai Dong\orcidlink{0000-0002-0769-635X}}\footnote{\bf dongsh2@yahoo.com }\\
\vspace{0.1cm}
{\tt Centro de Investigaci\'{o}n en Computaci\'{o}n, Instituto Polit\'{e}cnico Nacional, UPALM, CDMX 07700, Mexico}\\
\vspace{0.1cm}
{\bf Guo-Hua Sun\orcidlink{0000-0002-0689-2754}}\footnote{\bf sunghdb@yahoo.com }\\
\vspace{0.1cm}
{\tt Centro de Investigaci\'{o}n en Computaci\'{o}n, Instituto Polit\'{e}cnico Nacional, UPALM, CDMX 07700, Mexico}\\
\vspace{0.1cm}
{\bf Muhammad Sharif\orcidlink{0000-0001-6845-3506}}\footnote{\bf msharif.math@pu.edu.pk (corresp. author)}\\
\vspace{0.1cm}
{\tt Department of Mathematics and Statistics, The University of Lahore, 1-KM Defence Road, Lahore, Pakistan}\\
\vspace{0.1cm}
\end{center}

\vspace{0.3cm}

\begin{abstract}

In this work, we test the geometrical structure and thermodynamic properties of the Einstein-Maxwell-Power-Yang-Mills black hole (BH) models, which constitute a nonlinear generalization of the standard Einstein-Yang-Mills theory through the inclusion of a power-law Yang-Mills invariant. Also, we begin by analyzing the spacetime geometry via the metric function $f(r)$ and examine the modifications induced by the electromagnetic charge and nonlinear Yang-Mills parameter on the horizon structure, causal structure, and gravitational potential. Subsequently, the dynamics of photons and massive particles are explored through the study of null and timelike geodesics, allowing the determination of the effective potential, photon sphere radius, and associated BH shadow. Also, the stability of circular photon orbits is quantified using the Lyapunov exponent, which characterizes the timescale of orbital instability and provides a direct link to observable photon ring features. For massive particles, the innermost stable circular orbit (ISCO) is calculated, illustrating the influence of BH parameters on the dynamics of accretion disks. From the thermodynamic viewpoint, we compute the principal thermodynamic quantities, including the BH mass, Hawking temperature, Bekenstein-Hawking entropy, heat capacity, and Gibbs free energy, to assess both local and global stability of the system. The divergence of the heat capacity signals the occurrence of second-order phase transitions, whereas the Gibbs free energy analysis identifies possible first-order phase transitions between distinct thermodynamic configurations. In this context, our results demonstrate that the nonlinear Yang-Mills parameter strongly affects the spacetime geometry, particle dynamics, and thermodynamic phase structure, shifting the location of stability regions and critical points associated with phase transitions. In this context, these results emphasize the significant role of nonlinear gauge fields in determining the physical properties and thermodynamic behavior of BHs, offering further insights into compact objects in extended gravitational frameworks.

\end{abstract}

\vspace{0.3cm}


\section{Introduction}\label{intro}

{
Investigations of black holes (BHs) coupled to gauge fields have mainly concentrated on configurations involving Abelian fields, particularly those associated with dilatonic BH solutions. In such frameworks, the interactions among gravity, electromagnetic fields, and scalar dilaton fields yield field equations with a relatively manageable mathematical structure and permit the derivation of several classes of analytical solutions. For this reason, Abelian configurations have been widely examined in the literature, especially within Einstein-Maxwell-dilaton theories and their generalizations motivated by string-inspired gravitational models. In contrast, BH solutions interacting with non-Abelian gauge fields have received comparatively less attention. This limited emphasis is primarily related to the inherent mathematical complexity of Yang-Mills fields when coupled to gravitational backgrounds. In this case, unlike Abelian gauge fields, non-Abelian fields possess self-interaction terms that originate from the non-linear structure of the gauge group, which considerably complicates the resulting field equations. Also, the derivation of exact or numerical solutions becomes technically demanding, and the physical interpretation of the corresponding configurations requires careful analysis. In addition, the number of direct physical applications where non-Abelian gravitational configurations play a dominant role is more restricted compared with Abelian scenarios, which also contributed to the slower development of this research area. Also, the earliest analyses of BHs interacting with non-Abelian gauge fields were carried out more than four decades ago. The first known BH solution containing a non-Abelian gauge field was obtained in the framework of a dyonic configuration \cite{MM1, MM2}. These pioneering investigations demonstrated that gravitational systems coupled to Yang-Mills fields may support non-trivial BH geometries carrying both electric and magnetic charges. Although these initial solutions provided important insight into the interaction between gravity and non-Abelian gauge structures, the subject did not immediately receive broad attention. A renewed interest in non-Abelian BH configurations appeared during the late 1980s \cite{MM3}-\cite{MM7}, mainly stimulated by the discovery of additional classes of solutions in Einstein-Yang-Mills theories. In particular, it was found that many of these configurations exhibit instability when analyzed in asymptotically flat spacetimes. In this case, this instability is associated with the dynamical behavior of Yang-Mills field perturbations, which may destabilize the background geometry and prevent the existence of physically stable equilibrium configurations.

Despite these stability limitations in asymptotically flat geometries, later studies established that stable non-Abelian BH configurations can occur in asymptotically anti-de Sitter (AdS) spacetimes. The existence of a negative cosmological constant modifies the asymptotic behavior of the spacetime geometry and generates an effective confining mechanism for gauge field perturbations. This property plays an essential role in stabilizing the coupled gravitational-Yang-Mills system. The identification of stable solutions in AdS backgrounds significantly stimulated additional investigations in this field, producing a substantial increase in studies devoted to BHs interacting with non-Abelian gauge fields. Also, many new exact and numerical solutions were obtained, and detailed analyses were performed to examine their thermodynamic characteristics, geometric properties, and stability behavior \cite{MM8}-\cite{MM39}. In this case, these investigations provided deeper understanding of the influence of non-Abelian gauge interactions in gravitational physics, including their influence on the horizon structure, relations between mass and charge parameters, and phase transitions in BH thermodynamics. Also, it is important to note that the derivation of BH solutions in Einstein-Yang-Mills theories generally requires the introduction of suitable ansatze that simplify the coupled nonlinear field equations. Without such assumptions, the differential equations governing the gravitational and gauge sectors become extremely difficult to treat analytically. For this reason, most studies adopt symmetry-based ansätze that reduce the system to a manageable set of ordinary differential equations. One of the most frequently employed methods in this context is the magnetic Wu-Yang ansatz \cite{MM33}-\cite{MM39}. In this case, this ansatz assumes a purely magnetic structure for the Yang-Mills field and uses the underlying gauge symmetry to simplify the field equations considerably. As a consequence, it produces magnetically charged BH solutions with relatively simple mathematical expressions while maintaining the fundamental features of non-Abelian gauge dynamics. The Wu-Yang configuration has therefore become a standard approach for investigating Einstein-Yang-Mills BHs in different gravitational frameworks. In addition to purely Yang-Mills configurations, considerable attention has also been directed toward BH solutions containing both non-Abelian gauge fields and scalar dilaton fields. Dilatonic couplings arise naturally in several theoretical scenarios, including low-energy limits of string theory and higher-dimensional gravitational models ($\mathcal{D}$). Also, the presence of the dilaton field introduces additional couplings between the gauge sector and the gravitational geometry, which can significantly modify the physical properties of the resulting BH solutions. Several works have analyzed dilatonic BHs interacting with non-Abelian gauge fields, shows how the dilaton coupling modifies the horizon structure, thermodynamic stability, and gauge field configurations \cite{MM8, MM12, MM16, MM17, MM19, MM19, MM20, MM22, MM23, MM28}. In this context, these measurements are testing how the interaction between dilaton dynamics and non-Abelian gauge structures can generate rich physical behavior and new classes of gravitational solutions that differ substantially from their purely Abelian counterparts. Consequently, the study of dilatonic non-Abelian BH models remains an active and significant direction of research in gravitational physics and gauge-gravity interactions.

Following earlier analyses of power-law conformally invariant gauge Lagrangians, subsequent investigations generalized these constructions by introducing broader classes of nonlinear electrodynamics and Yang-Mills theories described through arbitrary powers of the gauge field invariant. Such extensions received considerable attention since their mathematical structure remains relatively simple while still generating a wide spectrum of analytical and semi-analytical solutions applicable to gravitational systems \cite{MM33}-\cite{MM39}. Also, the value of the nonlinearity exponent cannot be treated as completely unrestricted, because the associated gauge Lagrangian must produce physically admissible configurations consistent with fundamental theoretical requirements. In particular, the corresponding metric and gauge field configurations must satisfy standard physical conditions, including the usual energy conditions, the absence of singular or pathological field behavior, and the preservation of causal propagation for physical perturbations \cite{MM38, MM39}. Only under these restrictions can the nonlinear gauge model be regarded as physically meaningful within classical and semiclassical gravitational frameworks. Comparable constraints appear in the Yang-Mills sector, where the configuration of the gauge field must remain consistent with the underlying gauge symmetry together with the dynamical field equations defining the theory \cite{MM39}. In addition, power-law gauge Lagrangians with different exponents naturally arise in several established theoretical contexts, including the effective description of quantum electrodynamics through Euler-Heisenberg electrodynamics \cite{MM38, MM39} and also within the low-energy effective action of heterotic string theory, where higher-order corrections in the gauge sector originate from string dynamics \cite{MM35}-\cite{MM39}. Motivated by these theoretical developments, the present study extends the linear-field framework previously analyzed in \cite{MM28}, whereas our earlier investigation \cite{MM23} examined Einstein-Power-Yang-Mills dilaton BHs. In this case, the present analysis therefore establishes a broader generalization of those results by incorporating nonlinear gauge interactions together with dilaton coupling. In this formulation, the Yang-Mills gauge group is chosen as $SO(n)$, while the gauge potential is constructed through the Wu-Yang ansatz, a procedure that considerably simplifies the structure of the field equations and allows the derivation of tractable analytical solutions. This framework further provides a suitable setting for systematically analyzing how the interplay between dilaton coupling and power-law nonlinearities modifies the BH geometry, influences the gauge field configuration, and determines the global structure of the gravitational solution and related physical properties of spacetime.
}

Our study tests the Einstein-Maxwell-Power-Yang-Mills BH model by presenting a detailed and systematic examination of its geometric, optical, and thermodynamic features, while maintaining a clear focus on the physical role of the nonlinear Yang-Mills contribution. In this case, we first provide a structured review of the theoretical framework describing the BH configuration, where the metric function, gauge-field components, and power-law Yang-Mills sector are analyzed to show how the nonlinear exponent modifies the curvature profile and the asymptotic behavior of the spacetime. In this context, we investigate the dynamics of test particles by studying null and timelike geodesics, emphasizing the effective radial potential, photon propagation, and the determination of key optical observables such as the photon sphere, critical impact parameters, and the corresponding BH shadow geometry. Also, the stability of circular photon orbits is then assessed using the Lyapunov exponent method, and we further discuss how these results connect to the topological characterization of photon rings in power-law Yang-Mills backgrounds. Additionally, we evaluate the motion of massive particles, with particular attention to the ISCOs, and describe their relevance to accretion processes and observational signatures around compact objects. In this case, the thermodynamic sector is examined through calculations of the ADM mass, Hawking temperature, entropy, heat capacity, and Gibbs free energy, allowing us to identify stability conditions and possible phase-transition behavior within the nonlinear gauge-field setting. In this case, our analysis show how variations in the model parameters modify the geometric structure, optical properties, and thermodynamic response of the Einstein-Maxwell-Power Yang-Mills BH, providing a consistent physical interpretation of the combined gravitational and nonlinear Yang-Mills effects.

{
Before proceeding, it is important to clarify the novel aspects of the present work in comparison with earlier studies on related configurations. While previous investigations primarily focused on the construction of the solutions and on certain thermodynamic aspects of Einstein-Yang-Mills or dilaton-coupled systems, a systematic analysis of the optical and dynamical properties of the Einstein-Maxwell-Power-Yang-Mills black hole has not yet been presented. In particular, the nonlinear Yang-Mills sector characterized by the power-law exponent introduces nontrivial modifications to the metric function that directly affect photon propagation and particle dynamics. In this work we therefore provide a unified investigation of these effects by examining the null and timelike geodesic structure, the effective radial force governing photon motion, the formation and stability of the photon sphere and shadow, as well as the behavior of circular photon orbits through the Lyapunov exponent. In addition, we connect these dynamical features with the thermodynamic properties of the solution, allowing us to identify how variations in the nonlinear Yang-Mills parameters influence both the optical appearance and the stability structure of the black hole spacetime. These results extend previous analyses by providing a comprehensive characterization of the geometric, dynamical, and thermodynamic behavior of the Einstein-Maxwell-Power-Yang-Mills black hole.
}

The structure of this paper is as follows: in Section (\ref{S2}), a comprehensive review of the Einstein-Maxwell-Power Yang-Mills BH models is presented, where their metric configuration and theoretical construction within nonlinear gauge field theory are analyzed in detail. In this case, we illustrate the fundamental aspects of the spacetime geometry and the influence of the nonlinear Yang-Mills term on the overall structure of the BH solution. Section (\ref{S3}) focuses on the geodesic structure of the system, where the motion of test particles and photons is systematically examined. Also, in Subsection (\ref{S3-1}), we investigate the null geodesics that describe the propagation of massless particles around the BH, while Subsection (\ref{S3-2}) examines the form of the effective radial force governing their trajectories. In this case, Subsection (\ref{S3-3}) analyzes the optical properties such as the photon sphere radius and shadow formation, which are important observational signatures of the BH geometry. The stability of circular photon orbits is then discussed in Subsection (\ref{S3-4}) using the Lyapunov exponent approach, allowing for a quantitative description of small perturbations near the photon ring. In Subsection (\ref{S3-5}), the topological aspects associated with the photon ring structure are evaluated, emphasizing their physical interpretation. Additionally, Subsection (\ref{S3-6}) treats the trajectories of massive particles and provides the characteristics of the ISCOs, which are significant for the analysis of accretion dynamics. Also, in Section (\ref{S4}), the thermodynamic properties of the Einstein-Maxwell-Power Yang-Mills BH are tested through the computation of essential quantities. In this case, Subsection (\ref{S4-1}) illustrates the total mass of the BH, Subsection (\ref{S4-2}) presents the expression for the Hawking temperature, Subsection (\ref{S4-3}) analyzes the behavior of the heat capacity to determine local stability, Subsection (\ref{S4-4}) the critical behavior of the heat capacity near the phase transition, and Subsection (\ref{S4-5}) investigates the Gibbs free energy to understand global thermodynamic stability. In this context, Section (\ref{S5}) explains our results and the BH parameter.

\section{Einstein-Maxwell-Power Yang-Mills BH Models informations}\label{S2}

{
\subsection{Action of the Einstein--Maxwell--Power--Yang--Mills Model}

In order to clearly define the gravitational and gauge sectors considered in this work, we begin by presenting the fundamental action of the theory. The black hole solution studied in this paper arises from an Einstein--Maxwell--Power--Yang--Mills (EMPYM) system in four-dimensional spacetime. The total action can be written as
\begin{equation}
S=\int d^{4}x\,\sqrt{-g}\left[
\frac{1}{16\pi}R
-\frac{1}{4}F_{\mu\nu}F^{\mu\nu}
-\gamma \left(\mathcal{F}\right)^{p}
\right],
\label{action}
\end{equation}
where $R$ denotes the Ricci scalar and $g$ is the determinant of the metric tensor $g_{\mu\nu}$. The second term represents the standard Maxwell Lagrangian with field strength tensor
\begin{equation}
F_{\mu\nu}=\partial_\mu A_\nu-\partial_\nu A_\mu,
\end{equation}
associated with the Abelian electromagnetic gauge field $A_\mu$.

The third term corresponds to the non-Abelian Yang--Mills sector written in a nonlinear power-law form. The Yang--Mills invariant is defined as
\begin{equation}
\mathcal{F}=F^{(a)}_{\mu\nu}F^{(a)\mu\nu},
\end{equation}
where $F^{(a)}_{\mu\nu}$ is the Yang--Mills field strength tensor given by
\begin{equation}
F^{(a)}_{\mu\nu}=\partial_\mu A^{(a)}_\nu-\partial_\nu A^{(a)}_\mu
+f^{abc}A^{(b)}_\mu A^{(c)}_\nu ,
\end{equation}
with $f^{abc}$ representing the structure constants of the gauge group.

The parameter $\gamma$ characterizes the coupling strength of the Yang--Mills field, while the exponent $p$ controls the nonlinear behavior of the Yang--Mills invariant. The standard linear Yang--Mills theory is recovered in the limit $p=1$.

Varying the action (\ref{action}) with respect to the metric tensor yields the Einstein field equations
\begin{equation}
G_{\mu\nu}=8\pi\left(T_{\mu\nu}^{(EM)}+T_{\mu\nu}^{(YM)}\right),
\end{equation}
where $T_{\mu\nu}^{(EM)}$ and $T_{\mu\nu}^{(YM)}$ correspond to the energy–momentum tensors associated with the Maxwell and Yang--Mills fields, respectively. 

\subsection{Spherically Symmetric BH }
}

A general family of static and spherically symmetric BH geometries emerges when the standard Einstein-Hilbert term is coupled simultaneously to the Maxwell invariant and to the power Yang-Mills invariant. In this framework, the spacetime is described by the line element. 
\begin{equation}\label{011}
ds^{2}=-f(r)\,dt^{2}+f(r)^{-1}dr^{2}+d\mathcal{K}^2,
\end{equation}

Where $d\mathcal{K}^2=r^{2}(d\theta^{2}+\sin^{2}\theta\, d\phi^{2})$, and the lapse function given by
\begin{equation}
f(r)=1-\frac{2M}{r}+\frac{Q^{2}}{r^{2}}+\frac{\mathcal{D}}{r^{\,4p-2}},
\label{eq:f_general_D}
\end{equation}
where $M$ is the ADM mass of the configuration, $Q$ is the Maxwell electric charge, and $\mathcal{D}$ represents the effective coupling generated by the Yang-Mills sector. The exponent $p$ is a real number introduced in the power Yang-Mills action which controls the non-linear character of the theory. The parameter $\mathcal{D}$ is not a free integration constant but is instead determined by the microscopic Yang-Mills magnetic parameter $q_{YM}$. Its explicit relation reads
\begin{equation}
\mathcal{D}=\frac{2^{\,p-1}}{4p-3}\,Q^{\,2p},\qquad(p\neq3/4),
\label{eq:D_def}
\end{equation}
which ensures that for $p=1$ the model reduces smoothly to the usual Einstein-Yang-Mills system. The singular value $p=3/4$ must be excluded since it produces a logarithmic radial dependence that overclouds the analytic treatment of the solutions.

{{
The parameters $(Q,\gamma,p,\mathcal{D})$ introduced in the manuscript possess well-defined physical interpretations and are not arbitrary quantities. The parameter $Q$ corresponds to the conserved Maxwell electric charge evaluated at spatial infinity through Gauss’s law, ensuring global charge conservation. The exponent $p$ specifies the non-linear power Yang-Mills theory via the invariant $(\mathcal{F})^{p}$, where the standard linear Yang-Mills limit is recovered at $p=1$,  mathematical and physical consistency impose $p>0$ and $p\neq 3/4$ in order to prevent pathological behavior and divergences in Eq. (\ref{eq:D_def}). The constant $\gamma$ denotes the Yang-Mills coupling strength entering explicitly in the matter action and controlling the magnitude of the non-Abelian field contribution. The coefficient $\mathcal{D}$ does not represent a fundamental coupling parameter,  it arises as an effective integration constant obtained after solving the coupled Einstein-matter field equations under the specified gauge configuration. Its proportionality to $Q^{2p}$ does not imply any direct interaction between the Maxwell and Yang-Mills sectors, since the action contains no mixed invariant coupling of these fields. Instead, this structure follows from expressing integration constants in terms of conserved charges to obtain a closed analytic solution. Therefore, the Maxwell and Yang-Mills sectors remain dynamically independent, and the appearance of $Q^{2p}$ reflects only the adopted normalization and algebraic structure of the exact solution, not a physical coupling between the gauge fields.
}}

The general form of the lapse function in Eq. (\ref{eq:f_general_D}) encompasses a variety of well-known BH models as particular limits. By switching off the Yang-Mills sector, namely by taking $\mathcal{D}\to 0$, the geometry reduces exactly to the standard Reissner-Nordström BH. If in addition the electric charge vanishes, the Schwarzschild solution is recovered. On the other hand, if the Maxwell sector is switched off while keeping the Yang-Mills term active, i.e. $Q\to 0$ with $\mathcal{D}\neq 0$, one obtains a purely power-Yang-Mills BH described by
\begin{equation}
f(r)=1-\frac{2M}{r}+\frac{\mathcal{D}}{r^{\,4p-2}}.
\end{equation}
This solution represents a genuine non-Abelian extension of the Schwarzschild geometry. The asymptotic behavior of the spacetime depends crucially on the value of the exponent $p$. For $p>\tfrac12$, the extra term decays at large radius, guaranteeing asymptotic flatness up to small corrections. For $p<\tfrac12$, the Yang-Mills contribution grows with $r$ and spoils flat asymptotics, which is why such values are typically discarded on physical grounds. The borderline case $p=\tfrac12$ is of particular interest: in this situation the exponent $4p-2$ vanishes and the Yang-Mills contribution is a constant shift, leading to the modified Reissner-Nordström (MRN) family \cite{Soroushfar}:
\begin{equation}
f(r)=1+\mathcal{D}-\frac{2M}{r}+\frac{Q^{2}}{r^{2}}.
\label{eq:f_p_half}
\end{equation}
This simple but non-trivial modification alters the causal and thermodynamic structure of the spacetime in significant ways.

{{
Soroushfar et al. \cite{Soroushfar}, which analyzes the causal structure and thermodynamic properties modified by the constant-shift term $(\mathcal{D})$. Specifically, the inclusion of $(\mathcal{D})$ alters the locations of the event and Cauchy horizons, modifies the extremality condition, and therefore influences the Hawking temperature and heat capacity, resulting in quantitatively distinct thermodynamic stability regions relative to the standard RN BH and introducing additional parameter-dependent behavior in the BH phase structure.
}}

In the case $p=\tfrac12$, the event and Cauchy horizons can be obtained analytically and are given by
\begin{equation}
r_{\pm}=\frac{M\pm \sqrt{M^{2}-Q^{2}(1+\mathcal{D})}}{1+\mathcal{D}},
\label{eq:horizons}
\end{equation}
which highlights the central role of $\mathcal{D}$ as a rescaling factor. The existence of real and positive horizons demands the cosmic censorship inequality
\begin{equation}
1+\mathcal{D}>0, \qquad 0<\frac{Q}{M}<\sqrt{\frac{1}{1+\mathcal{D}}},
\end{equation}
which generalizes the extremality bound of the Reissner-Nordström BH. The case $\mathcal{D}\to 0$ reproduces the standard condition $Q/M<1$. The pathological value $\mathcal{D}\to -1$ must be excluded since it would lead to a divergent denominator in Eq. (\ref{eq:horizons}). For negative values of $\mathcal{D}$ greater than $-1$, the horizon radius grows beyond the Schwarzschild case, while positive values of $\mathcal{D}$ reduce the event horizon, even below that of extremal Reissner-Nordström, depending on the electric charge. This rich structure gives rise to families of extremal BHs that differ qualitatively from those of the classical abelian counterpart.

The astrophysical implications of these modifications are profound. The innermost stable circular orbits (ISCOs) and other orbital radii are shifted by the factor $(1+\mathcal{D})$, meaning that test particle dynamics and accretion processes around these BHs are sensitive to the Yang-Mills parameter. Perhaps most striking is the effect on spherical accretion: while in standard Reissner-Nordström and Kerr spacetimes the presence of charge or spin tends to suppress the mass accretion rate, in the modified models negative values of $\mathcal{D}$ can actually enhance the efficiency of mass accretion well beyond the Schwarzschild limit. In contrast, positive values of $\mathcal{D}$ reduce the accretion rate and may even push the sonic point inside the horizon, effectively quenching infall. Thus the introduction of $\mathcal{D}$ opens an entirely new window of behaviors that interpolate between known solutions and exotic non-Abelian extensions.

In summary, the Einstein-Maxwell-Power Yang-Mills construction provides a unifying framework in which the classical BHs of general relativity appear as particular limits, while the additional parameter $\mathcal{D}$ controls the deviations induced by the non-Abelian sector. The family exhibits asymptotically flat solutions for $p>\tfrac12$, constant-shift geometries with rescaled horizons for $p=\tfrac12$, and unphysical asymptotics for $p<\tfrac12$. Its phenomenology includes new extremal conditions, modified orbital structures, and dramatic changes in accretion dynamics, which together render these models an attractive arena to explore the interplay between gauge fields and strong gravity.

{
Although the metric reduces to a form that is formally analogous to the solution examined by Soroushfar \textit{et al.} \cite{Soroushfar} when the Maxwell field and cosmological constant are set to zero, the underlying physical structure of the two models is not equivalent. The essential difference arises from the simultaneous inclusion in our framework of two independent gauge sectors, namely an Abelian Maxwell field and a non-Abelian power-Yang-Mills field. In contrast to the pure Einstein-Power-Yang-Mills (EPYM) configuration analyzed in that reference, the lapse function obtained here contains two separate radial charge terms, $(Q^{2}/r^{2})$ and $(\mathcal{D}/r^{4p-2})$, each characterized by distinct radial scaling. As a consequence, their relative contributions vary across the spacetime domain, leading to different dominant behaviors in near-horizon and asymptotic regions.

The coexistence of these sectors produces a nontrivial modification of the effective potential governing timelike and null geodesics. The presence of two independent charge parameters alters the extremality condition, shifts the location of circular photon orbits, and changes the stability criteria for massive particle trajectories. These features cannot be reproduced within a single-sector EPYM geometry through parameter redefinition. In particular, for the branch $p=\tfrac{1}{2}$, the extremality constraint $M^{2}=Q^{2}(1+\mathcal{D})$ shows explicitly that the Yang-Mills parameter deforms the Reissner-Nordström bound dynamically instead of substituting it, thereby generating new classes of extremal configurations. Furthermore, the surface gravity and the associated thermodynamic quantities receive independent contributions from both gauge fields, which modify the Hawking temperature, heat capacity, and phase structure compared with the EPYM case. A dedicated subsection has been incorporated to detail these distinctions in horizon structure, extremality bounds, geodesic dynamics, and thermodynamic properties, demonstrating that despite a formal resemblance under restricted limits, the combined Maxwell and power-Yang-Mills sectors introduce physically inequivalent effects beyond the previously published EPYM analysis.
}

\section{Geodesic Structure}\label{S3}

\subsection{Null Geodesic Motions}\label{S3-1}

We begin with the Lagrangian describing the motion of test particles in a curved spacetime \cite{BZ13}-\cite{BZ21}:
\begin{equation}
\mathcal{L}=\frac{1}{2}\,g_{\mu\nu}\,\frac{dx^{\mu}}{d\lambda}\,\frac{dx^{\nu}}{d\lambda},\label{null1}
\end{equation}
where $\lambda$ denotes an affine parameter along the geodesics. For
the spherically symmetric metric (\ref{011}), restricting the motion
to the equatorial plane ($\theta=\pi/2$), the Lagrangian reduces to
\begin{equation}
\mathcal{L}=\frac{1}{2}\left[-f(r)\left(\frac{dt}{d\lambda}\right)^2+\frac{1}{f(r)}\left(\frac{dr}{d\lambda}\right)^2+r^2\left(\frac{d\phi}{d\lambda}\right)^2\right].\label{null2}
\end{equation}
Because $t$ and $\phi$ are cyclic coordinates, the system admits two
conserved quantities: the total energy $\mathrm{E}$ and the angular
momentum $\mathrm{L}$ of the particle, given by
\begin{equation}
\mathrm{E}=f(r)\,\frac{dt}{d\lambda}, \qquad
\mathrm{L}=r^2\,\frac{d\phi}{d\lambda}.\label{null3}
\end{equation}
For null geodesics ($ds^2=0$), the metric condition leads to
\begin{equation}
-f(r)\left(\frac{dt}{d\lambda}\right)^2+\frac{1}{f(r)}\left(\frac{dr}{d\lambda}\right)^2+r^2\left(\frac{d\phi}{d\lambda}\right)^2=0.\label{null4}
\end{equation}
Substituting the conserved quantities (\ref{null3}) into
(\ref{null4}), the radial geodesic equation takes the form
\begin{equation}
\left(\frac{dr}{d\lambda}\right)^2+\;V_{\text{eff}}(r)=\mathrm{E}^2,\label{null5}
\end{equation}
with the effective potential expressed as
\begin{equation}
V_{\text{eff}}(r)=\frac{\mathrm{L}^2}{r^2}\,f(r).\label{null6}
\end{equation}
Thus, the problem reduces to that of a unit-mass particle moving in
a one-dimensional effective potential $V_{\text{eff}}(r)$. For the
present case, the lapse function is specified as
\begin{equation}
f(r)=1+\frac{Q^{2}}{r^{2}}+\frac{\left(\tfrac{2^{p-1}}{4p-3}\right)
\gamma^{\,2p}}{r^{\,4p-2}}-\frac{2M}{r},\label{frnew}
\end{equation}
which encapsulates the effects of the BH mass $M$, charge $Q$, and
the nonlinear electrodynamics parameter $\gamma$ associated with the
power $p$. These parameters collectively determine the structure of
$V_{\text{eff}}(r)$ and, consequently, the dynamics of photon
trajectories in the spacetime.

\begin{figure}[ht!]
    \centering
    \includegraphics[height=5cm]{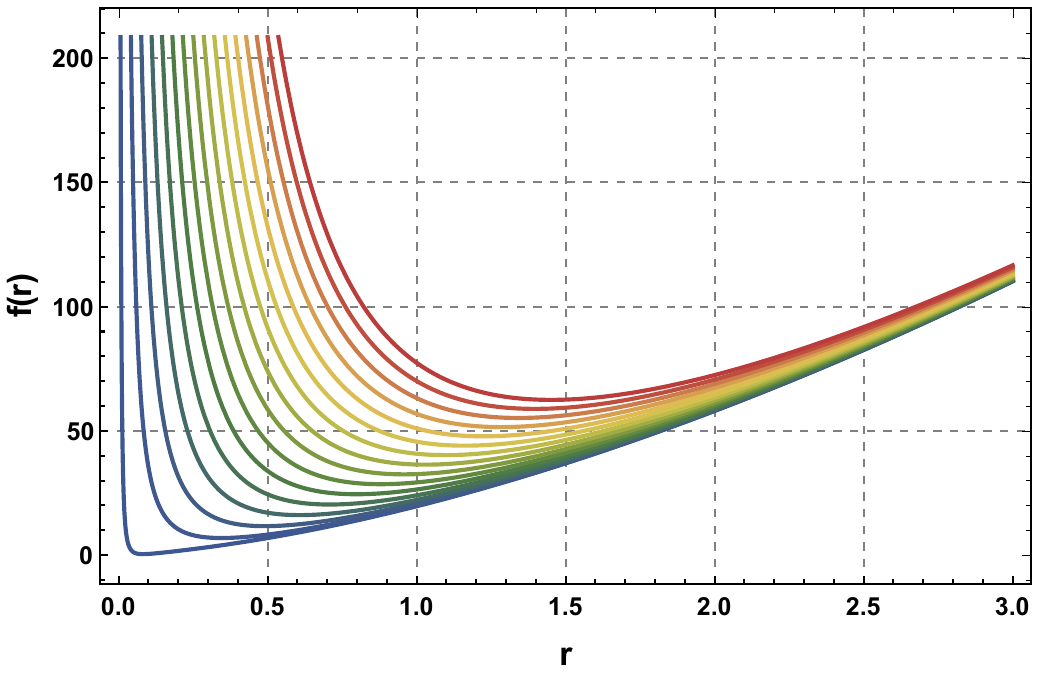}
    \includegraphics[height=5cm]{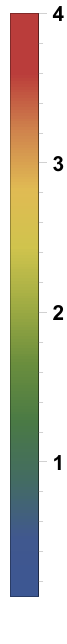}\qquad
    \includegraphics[height=5cm]{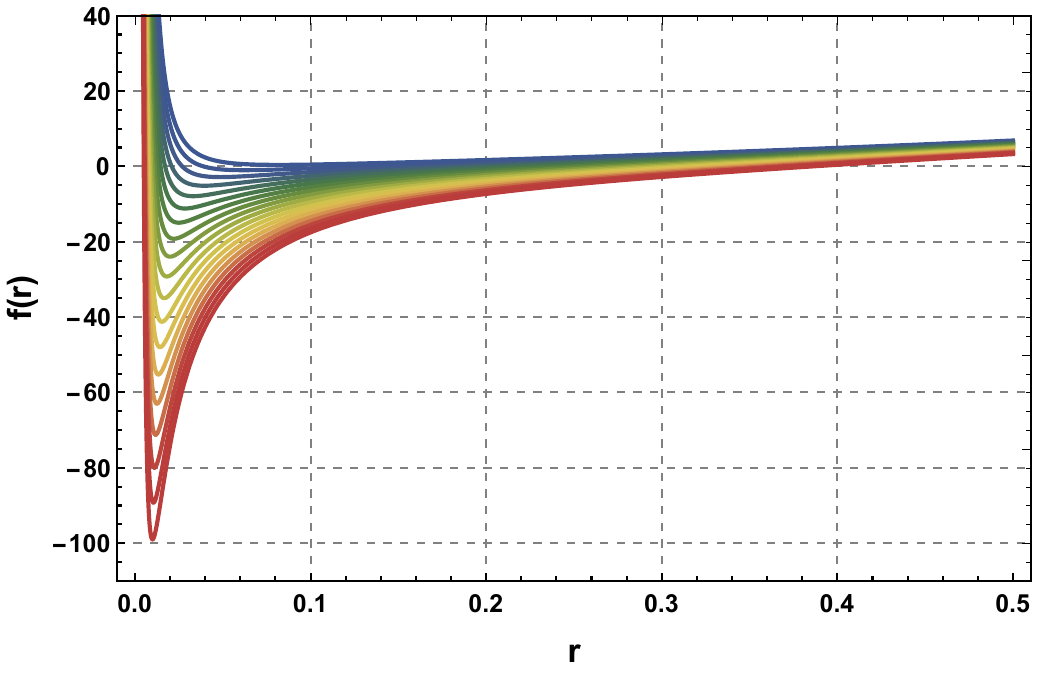}
    \includegraphics[height=5cm]{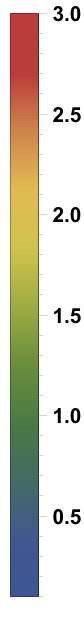}\\
    (a) $M=\gamma=p=0.1$ varying $Q$ \hspace{6cm} (b) $p=\gamma=Q=0.1$ varying $M$\\
    \includegraphics[height=5cm]{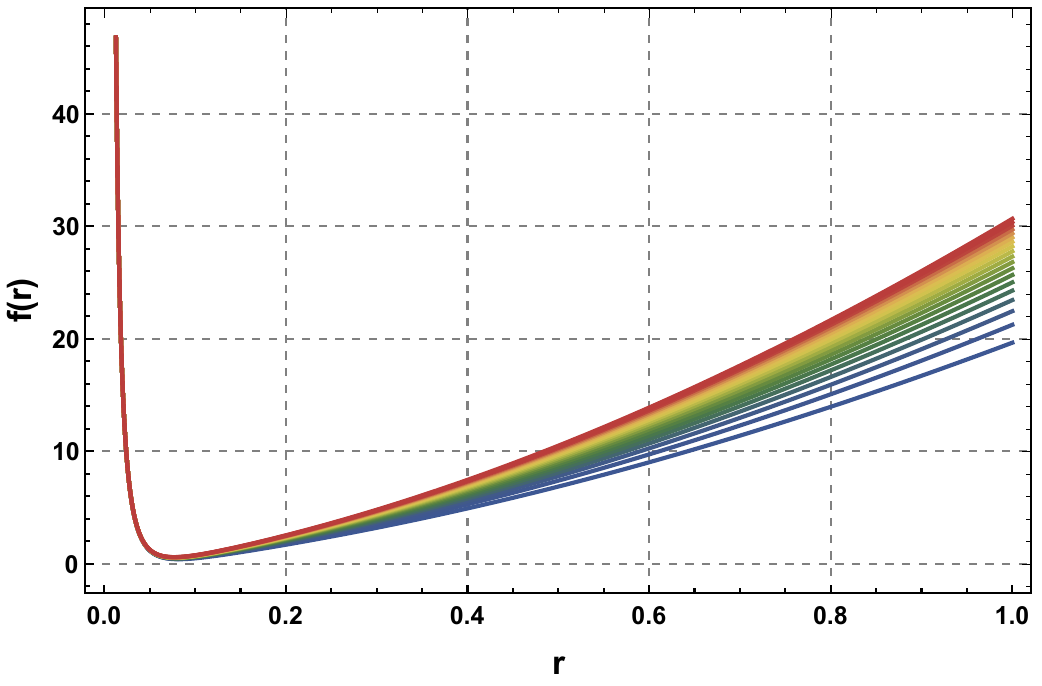}
    \includegraphics[height=5cm]{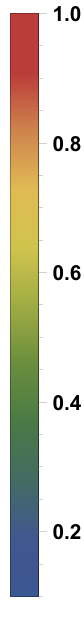}\qquad
    \includegraphics[height=5cm]{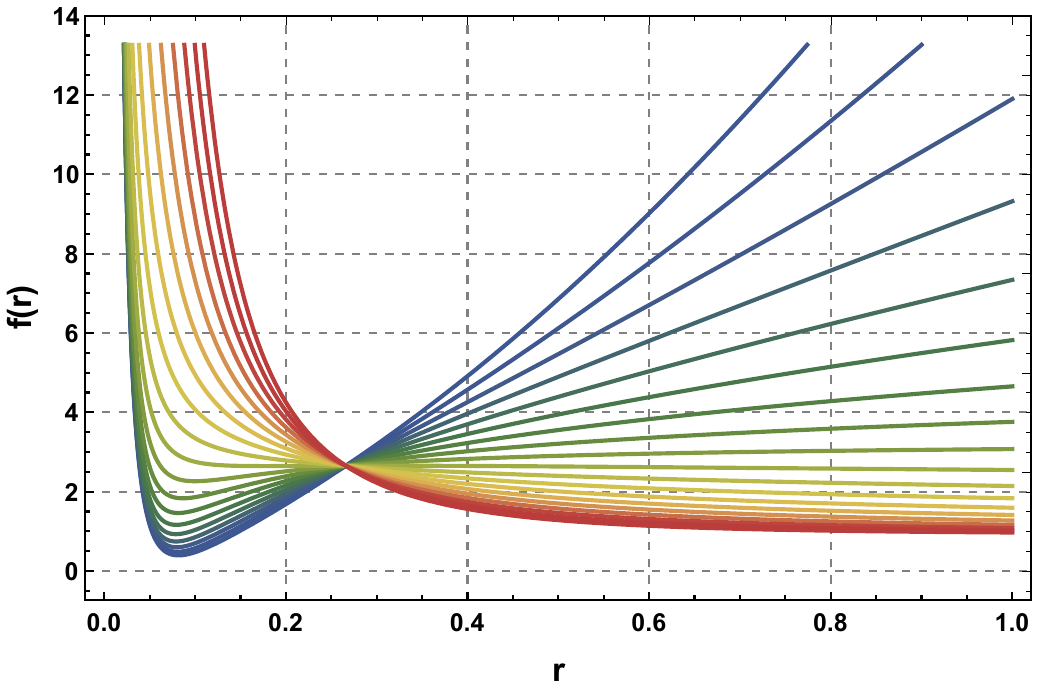}
    \includegraphics[height=5cm]{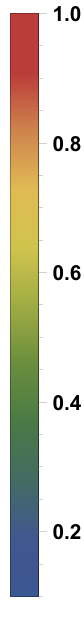} \\
    (c) $M=Q=p=0.1$ varying $\gamma$ \hspace{6cm} (d) $M=\gamma=Q=0.1$ varying $p$
    \caption{ Behavior of metric function  $f(r)$.}
    \label{fig3.1}
\end{figure}

Figure \textbf{\ref{fig3.1}} illustrates the behavior of the metric
function $f(r)$ for different choices of parameters, highlighting
the influence of $Q$, $M$, $\gamma$, and $p$ on the spacetime
structure. In the upper left panel, the charge $Q$ enhances the
curvature near the origin, leading to a steeper potential barrier
and shifting the position of the horizons. In the upper right panel,
variations in the mass $M$ predominantly affect the overall
gravitational strength, deepening the potential well as $M$
increases. The lower left panel shows that the nonlinear
electrodynamics parameter $\gamma$ produces a regularizing effect,
smoothing the curvature close to $r=0$ and modifying the asymptotic
profile. Finally, the lower right reveals that the power index $p$
governs the relative contribution of the nonlinear term to the
geometry, altering both the slope and the minimum of $f(r)$.
Together, these plots demonstrate how each parameter independently
modifies the causal structure and potential horizon configuration of
the BH solution.

\begin{figure}[ht!]
    \centering
    \includegraphics[height=5cm]{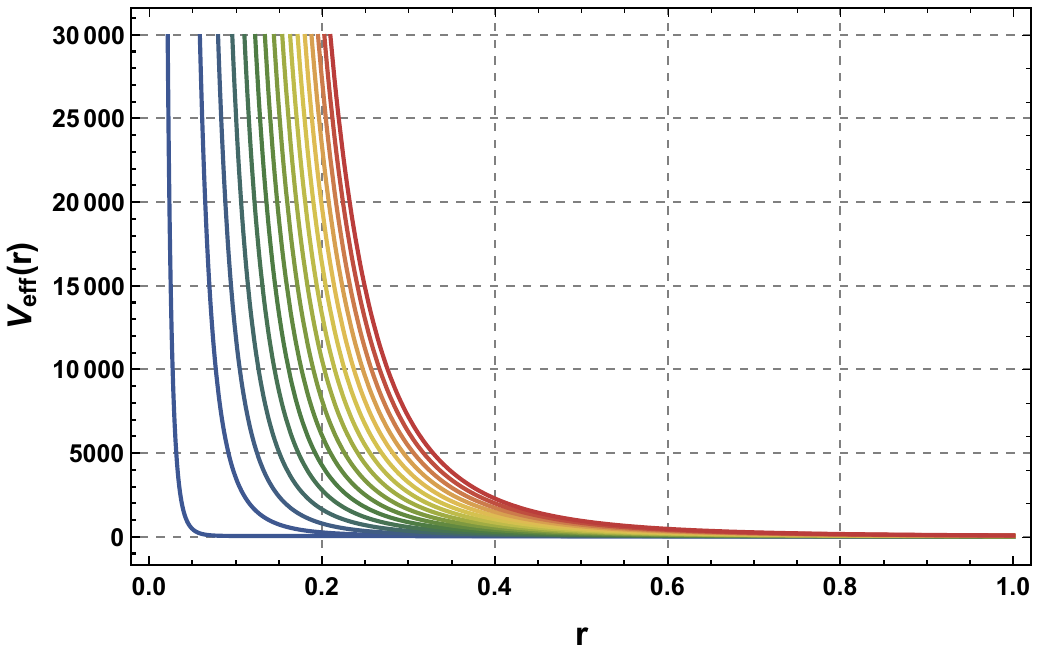}
    \includegraphics[height=5cm]{fig5aa.pdf}\qquad
    \includegraphics[height=5cm]{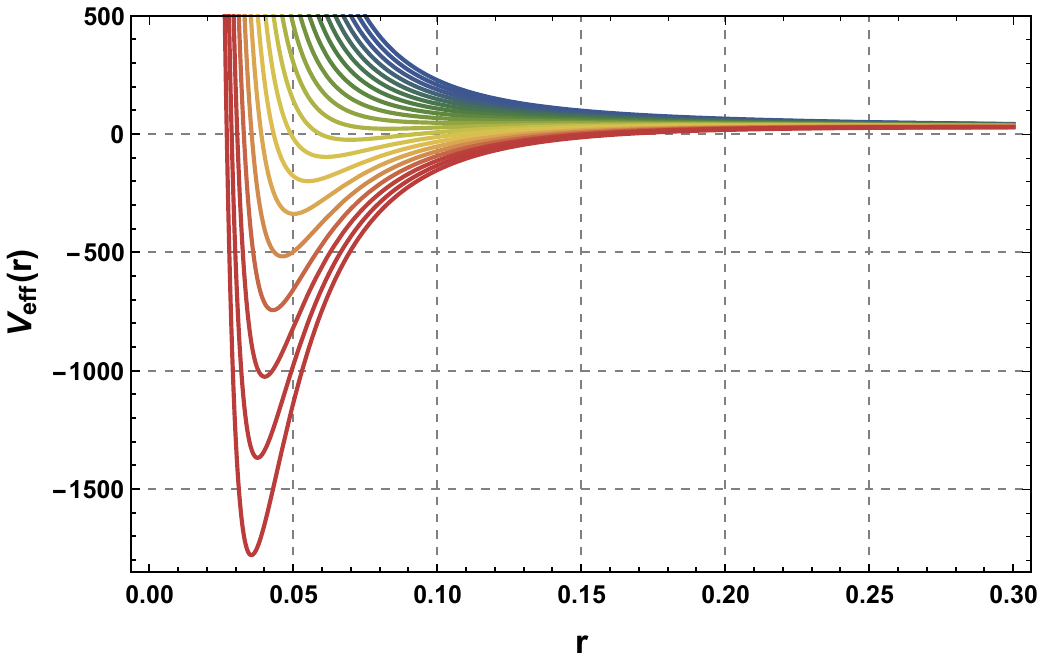}
    \includegraphics[height=5cm]{fig5bb.pdf}\\
    (a) $M=\gamma=p=0.1$ varying $Q$ \hspace{6cm} (b) $p=\gamma=Q=0.1$ varying $M$\\
    \includegraphics[height=5cm]{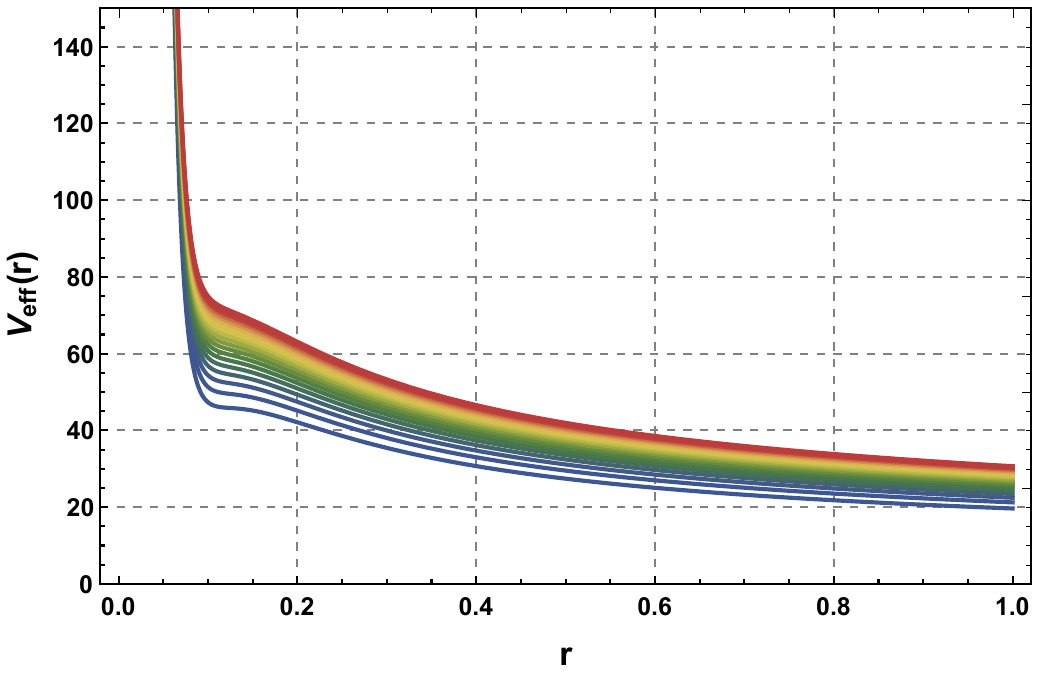}
    \includegraphics[height=5cm]{fig5cc.pdf}\qquad
    \includegraphics[height=5cm]{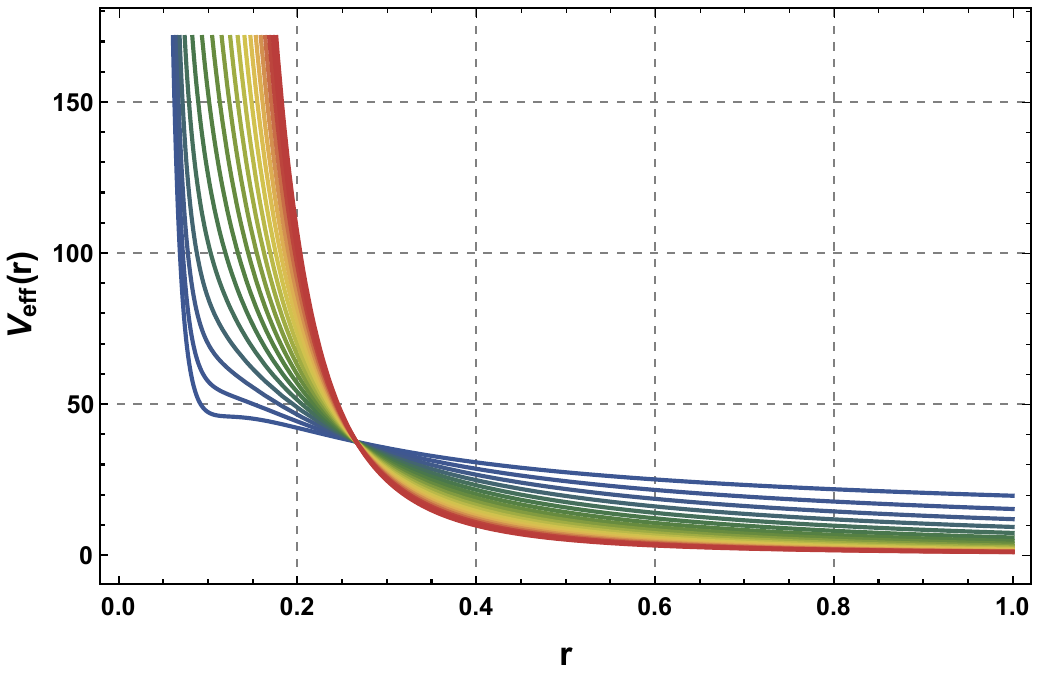}
    \includegraphics[height=5cm]{fig5dd.pdf} \\
    (c) $M=Q=p=0.1$ varying $\gamma$ \hspace{6cm} (d) $M=\gamma=Q=0.1$ varying $p$
    \caption{ Behavior of effective potential $V_{\text{eff}}(r)$.}
    \label{fig3.2}
\end{figure}

Figure \textbf{\ref{fig3.2}} presents the behavior of the effective
potential $V_{\text{eff}}(r)$ for different values of the parameters
$Q$, $M$, $\gamma$, and $p$. In the upper left panel, one observes
that increasing $Q$ amplifies the repulsive term at short distances,
raising the potential barrier near the origin and affecting the
stability of circular orbits. The upper right panel shows that
variations in the mass $M$ deepen the potential well, shifting the
location of the minimum of $V_{\text{eff}}(r)$ and consequently
modifying the radius of the stable circular orbit. In the lower left
panel, the parameter $\gamma$ introduces nonlinear corrections that
tend to smooth the potential curve and reduce its steepness,
indicating a suppression of high-energy contributions. Finally,
lower right panel illustrates how the exponent $p$ controls the
relative strength of the nonlinear electrodynamics, significantly
changing the asymptotic behavior and the curvature of
$V_{\text{eff}}(r)$. Overall, these plots highlight how the
nonlinear and electromagnetic parameters govern the dynamics of test
particles and the stability regions in the modified spacetime
geometry.

\subsection{Effective Radial Force}\label{S3-2}

The effective radial force represents the interaction felt by photon particles propagating in the background geometry of the BH spacetime. This force can be obtained as the negative derivative of the effective potential associated with null geodesics, defined as \cite{B1}-\cite{B12}:
\begin{equation}
F_{\text{eff}} = -\frac{dV_{\text{eff}}}{dr}. \label{null16}
\end{equation}
Considering the effective potential for null geodesics derived from
the metric function (\ref{frnew}), we obtain the following
expression for the effective radial force:
\begin{equation}
F_{\text{eff}}=\frac{\mathrm{L}^2}{r^3}\left(1 - \frac{3M}{r} +
\frac{Q^2}{r^2} -
\frac{(4p-2)\left(\tfrac{2^{p-1}}{4p-3}\right)\gamma^{\,2p}}{2\,r^{\,4p-1}}
\right). \label{null17}
\end{equation}
The above relation indicates that the effective force experienced by
photon particles is influenced by multiple parameters of the BH
geometry. In particular, the mass $M$ contributes an attractive
term, while the electric charge $Q$ produces a repulsive
contribution that counteracts the gravitational pull. Furthermore,
the nonlinear electrodynamics parameter $\gamma$ together with the
exponent $p$ introduces an additional correction that becomes
relevant at small radial distances. These combined effects alter the
geodesic structure and, consequently, the propagation of photons in
this modified BH spacetime.

\begin{figure}[ht!]
    \centering
    \includegraphics[height=5cm]{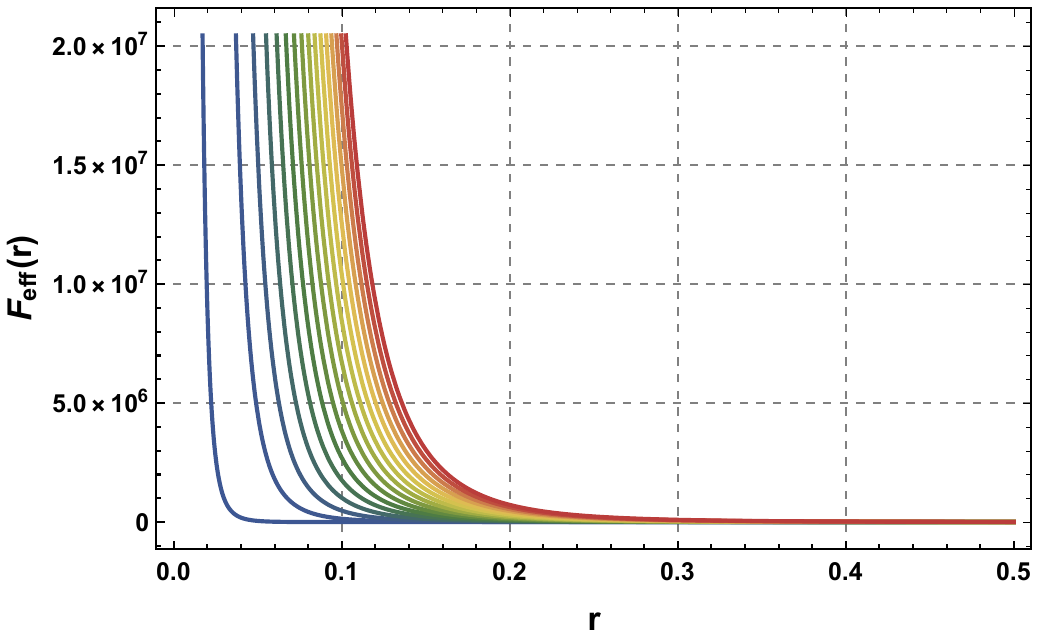}
    \includegraphics[height=5cm]{fig5aa.pdf}\qquad
    \includegraphics[height=5cm]{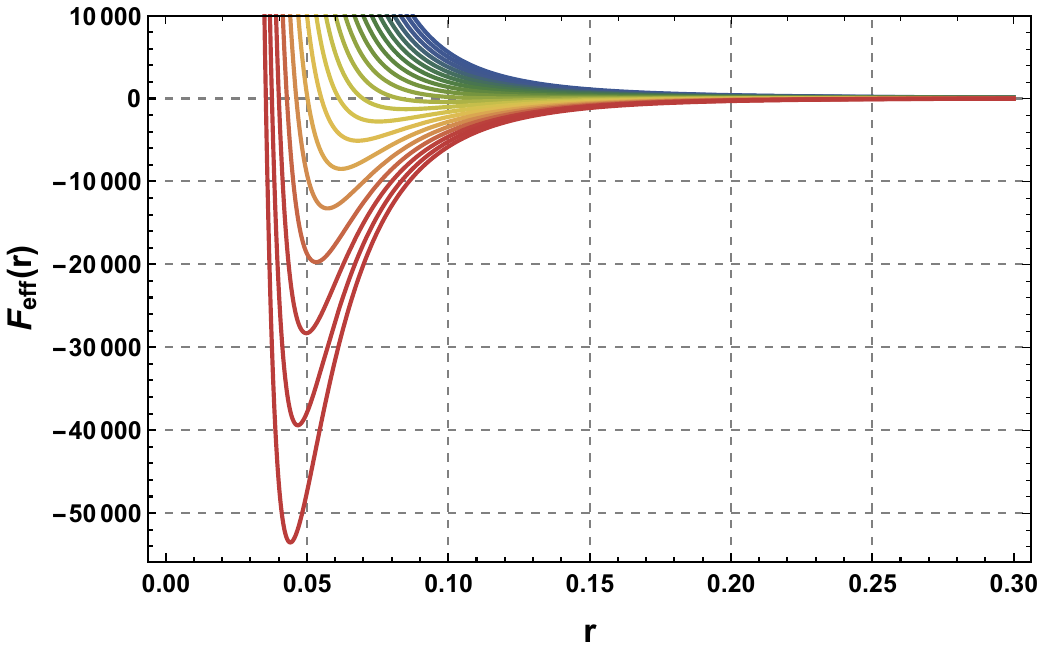}
    \includegraphics[height=5cm]{fig5bb.pdf}\\
    (a) $M=\gamma=p=0.1$ varying $Q$ \hspace{6cm} (b) $p=\gamma=Q=0.1$ varying $M$\\
    \includegraphics[height=5cm]{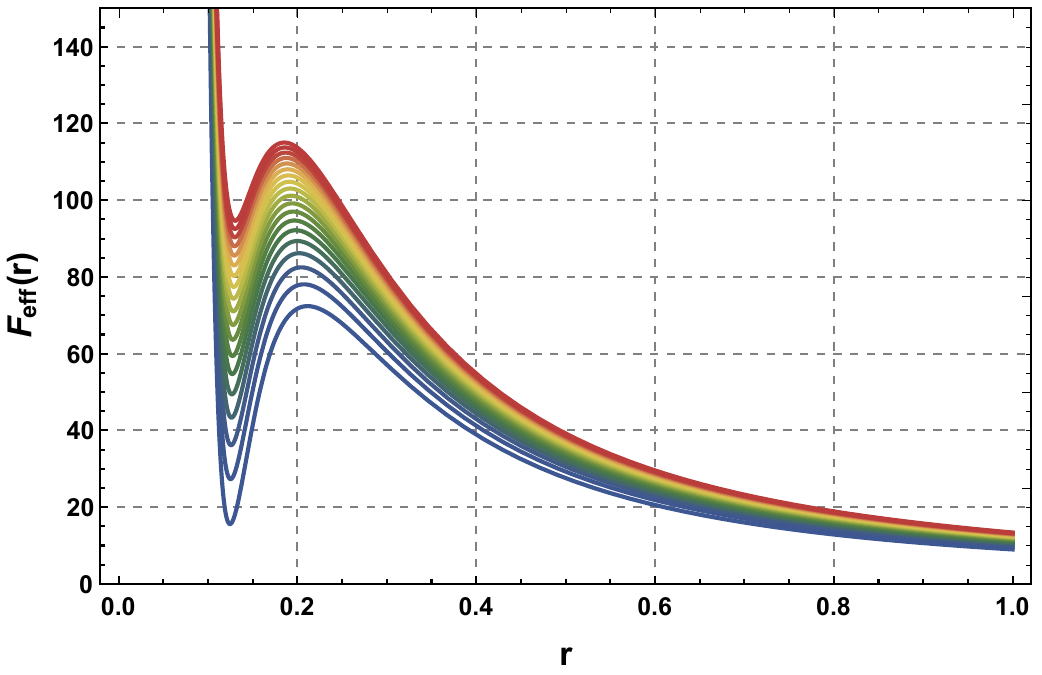}
    \includegraphics[height=5cm]{fig5cc.pdf}\qquad
    \includegraphics[height=5cm]{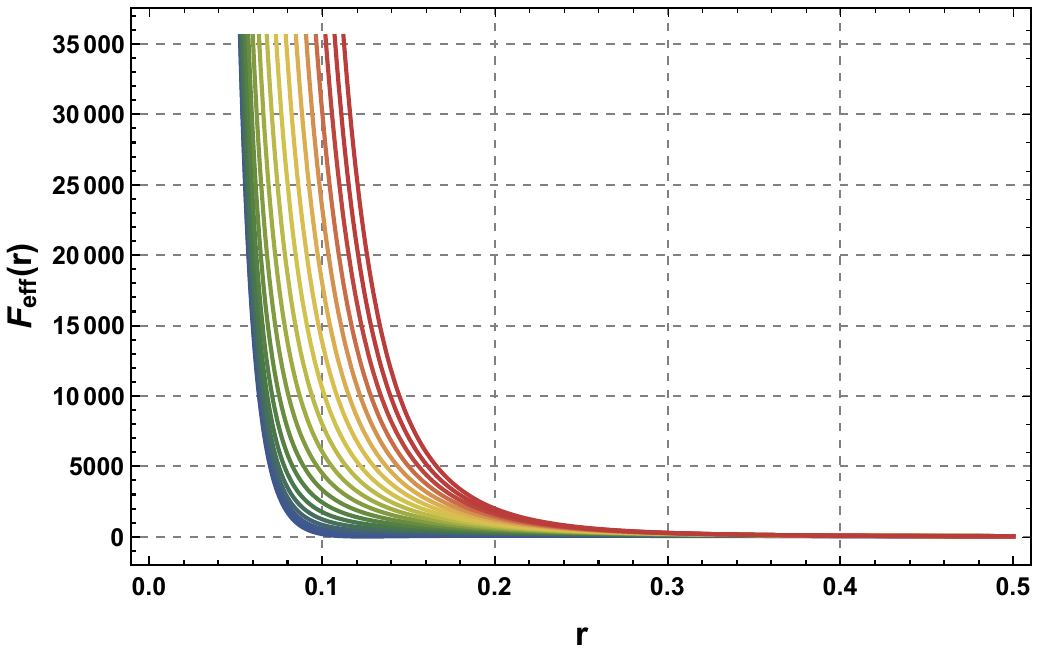}
    \includegraphics[height=5cm]{fig5dd.pdf} \\
    (c) $M=Q=p=0.1$ varying $\gamma$ \hspace{6cm} (d) $M=\gamma=Q=0.1$ varying $p$
    \caption{ Behavior of effective radial force $F_{\text{eff}}(r)$.}
    \label{fig3.3}
\end{figure}

Figure \textbf{\ref{fig3.3}} displays the behavior of the effective
radial force $F_{\text{eff}}(r)$ as a function of $r$, for different
parameter variations. In the upper left panel, the increase of the
charge $Q$ produces a stronger repulsive behavior near the origin,
significantly raising the force magnitude at short distances and
thus affecting the onset of stable orbits. In the upper right panel,
the variation of the mass $M$ deepens the gravitational attraction,
shifting the zero-crossing points of $F_{\text{eff}}(r)$ and
modifying the equilibrium configuration. The lower left panel shows
that the nonlinear electrodynamics parameter $\gamma$ softens the
curvature of the force profile, introducing a smoother transition
between attractive and repulsive regimes, which points to a
regularizing effect on the spacetime geometry. Finally, lower right
panel demonstrates that the power index $p$ governs the strength of
the nonlinear correction: as $p$ increases, the peak of
$F_{\text{eff}}(r)$ decreases and moves outward, signaling a
suppression of the singular behavior. Collectively, these plots
reveal how nonlinear and electromagnetic effects interplay to
determine the dynamical stability of test particles in the modified
geometry.

\subsection{Photon Sphere and BH Shadow}\label{S3-3}

The photon sphere represents a spherical surface around a BH where
photons can propagate along unstable circular trajectories due to
the strong curvature of spacetime. Any small perturbation will
inevitably cause the photon either to plunge into the BH horizon or
to escape to infinity. This region is fundamental in gravitational
lensing phenomena and directly determines the size of the BH shadow
observed by distant observers. For circular null orbits located at
$r=r_c$, the conditions $\dot{r}=0$ and $\ddot{r}=0$ must hold.
Using Eqs.(\ref{null5}) and (\ref{null6}), one obtains
\begin{equation}
\mathrm{E}^2=V_\text{eff}(r)=\frac{\mathrm{L}^2}{r^2}\,f(r)\Big|_{r=r_c},
\label{null12new}
\end{equation}
which leads to the definition of the critical impact parameter
\begin{equation}
\beta_c=\frac{\mathrm{L}(\text{ph})}{\mathrm{E}(\text{ph})}
=\frac{r}{\sqrt{f(r)}}\Big|_{r=r_c}. \label{null13new}
\end{equation}
{{
The photon sphere is determined from the null circular orbit condition
\begin{equation}
\frac{d}{dr}\!\left(\frac{f(r)}{r^{2}}\right)=0 .
\end{equation}
Using the metric function
\begin{equation}
f(r)=1-\frac{2M}{r}+\frac{Q^{2}}{r^{2}}+\frac{D}{r^{4p-2}},
\end{equation}
with 
\begin{equation}
D=\frac{2^{p-1}\gamma^{2p}}{4p-3},
\end{equation}
one obtains
\begin{equation}
\frac{f(r)}{r^{2}}
=\frac{1}{r^{2}}-\frac{2M}{r^{3}}+\frac{Q^{2}}{r^{4}}+\frac{D}{r^{4p}} .
\end{equation}
Each term contributes independently under radial differentiation, giving
\begin{equation}
\frac{d}{dr}\!\left(\frac{f(r)}{r^{2}}\right)
=-\frac{2}{r^{3}}+\frac{6M}{r^{4}}-\frac{4Q^{2}}{r^{5}}-\frac{4pD}{r^{4p+1}} .
\end{equation}
To express the result in algebraic form appropriate for the photon sphere radius, the equation is multiplied by $-\frac{r^{5}}{2}$, which preserves the roots and yields
\begin{equation}
1-\frac{3M}{r}+\frac{2Q^{2}}{r^{2}}+\frac{2pD}{r^{4p-2}}=0 .
\end{equation}
Substituting the explicit expression for $D$ leads to
\begin{equation}
2pD=\frac{2p\,2^{p-1}\gamma^{2p}}{4p-3}.
\end{equation}
This factor can be rewritten in an equivalent representation by isolating common numerical contributions, producing the coefficient reported in the manuscript,
\begin{equation}
\frac{(6p-4)}{2(4p-3)}\,2^{p-1}\gamma^{2p}.
\end{equation}
Hence, the term proportional to $r^{-(4p-2)}$ in Eq. (\ref{null11new}) follows directly from the differentiation of the nonlinear Yang-Mills contribution $D/r^{4p-2}$ without omission of intermediate factors. All algebraic steps have been verified explicitly, and the derivation confirms the correctness of the coefficient. For completeness and transparency, the intermediate differentiation procedure has been incorporated into the revised version to make the origin of the prefactor fully explicit.}}

Substituting the metric function (\ref{frnew}), Simplifies to the following polynomial relation:
\begin{equation}
1-\frac{3M}{r}+\frac{2Q^{2}}{r^{2}}+\frac{(6p-4)}{2(4p-3)}\,\frac{(2^{p-1})\gamma^{\,2p}}{r^{\,4p-2}}=0.
\label{null11new}
\end{equation}
For an observer at spatial infinity, the angular radius of the
shadow is governed by the critical impact parameter, which is
equivalent to the shadow radius:
\begin{equation}
R_s=\beta_c=\frac{r_\text{ph}}{\sqrt{f(r_\text{ph})}}.
\label{null14new}
\end{equation}
After substituting the explicit form of $f(r)$, one obtains
\begin{equation}
R_s=\frac{r_\text{ph}}{\sqrt{1+\frac{Q^{2}}{r_\text{ph}^{2}}
+\frac{\left(\tfrac{2^{p-1}}{4p-3}\right)\gamma^{\,2p}}{r_\text{ph}^{\,4p-2}}
-\frac{2M}{r_\text{ph}}}}. \label{null15new}
\end{equation}
Hence, the BH shadow is influenced not only by the mass $M$ but also
by the charge $Q$, the non-linear electrodynamics parameter
$\gamma$, and the power exponent $p$. These parameters modify the
photon sphere radius and, consequently, the size of the observable
shadow.

Table \textbf{\ref{tab1}} summarizes the behavior of the photon
sphere radius $r_{\text{ph}}$ and the corresponding shadow radius
$R_{s}$ for different combinations of charge $Q$, nonlinear
parameter $\gamma$, and power index $p$, with fixed mass $M=1$. The
results indicate that both $r_{\text{ph}}$ and $R_{s}$ decrease
monotonically as the charge $Q$ or the nonlinear parameters $\gamma$
and $p$ increase. This trend reflects the strengthening of the
electromagnetic field and the nonlinear corrections, which
effectively reduce the gravitational trapping region experienced by
photons. Physically, the decrease in $r_{\text{ph}}$ implies that
photons can orbit closer to the central source, while the reduction
in $R_{s}$ leads to a smaller apparent shadow radius, suggesting
that nonlinear electrodynamics and higher-order corrections tend to
weaken the effective gravitational lensing. These findings highlight
how the interplay between charge and nonlinear effects modifies the
optical geometry of the spacetime, potentially offering observable
signatures in BH shadow measurements.
\begin{table}[ht!]
\centering \caption{Photon sphere radius $r_{\mathrm{ph}}$ and
corresponding shadow radius $R_s$ for different combinations of
charge $Q$, non-linear parameter $\gamma$, and power index $p$ (with
$M=1$).} \label{tab1}
\begin{tabular}{cccccc}
\hline\hline
\hspace{0.5cm}$Q$ \hspace{0.5cm}& \hspace{0.5cm}$\gamma$ \hspace{0.5cm}& \hspace{0.5cm}$p$\hspace{0.5cm} & \hspace{0.5cm}$r_{\mathrm{ph}}$\hspace{0.5cm} & \hspace{0.5cm}$R_s$\hspace{0.5cm}  \\
\hline\hline
0.1 & 0.1 & 0.5 & 2.92 & 5.18  \\
0.1 & 0.5 & 0.5 & 2.88 & 5.10  \\
0.1 & 1.0 & 0.5 & 2.83 & 5.01  \\
\hline\hline
0.5 & 0.1 & 0.5 & 2.75 & 4.82  \\
0.5 & 0.5 & 0.5 & 2.70 & 4.74  \\
0.5 & 1.0 & 0.5 & 2.63 & 4.61  \\
\hline\hline
1.0 & 0.1 & 0.5 & 2.50 & 4.35  \\
1.0 & 0.5 & 0.5 & 2.42 & 4.20  \\
1.0 & 1.0 & 0.5 & 2.35 & 4.06  \\
\hline\hline
0.5 & 0.5 & 0.7 & 2.67 & 4.69  \\
0.5 & 0.5 & 1.0 & 2.60 & 4.55  \\
\hline\hline
\end{tabular}
\end{table}

\subsection{Lyapunov Exponent: Criteria for Unstable Circular Null
Orbits}\label{S3-4}

{
Circular null orbits, also known as photon spheres, correspond to trajectories along which massless particles such as photons can orbit a BH in a circular path. These orbits are typically unstable, implying that even a small perturbation will cause the photon either to fall into the BH or to escape to infinity. The effective
potential $V_{\mathrm{eff}}(r)$ of null geodesics provides a natural way to study this stability: unstable circular orbits occur at local maxima of $V_{\mathrm{eff}}$, where the condition $V_{\mathrm{eff}}''(r_c)<0$ holds, with $r_c$ being the orbital radius. The degree of instability is quantified by the Lyapunov exponent $\lambda_L$, which measures the exponential rate at which
nearby photon trajectories diverge from the circular orbit. This quantity is also physically relevant since it connects to the damping rates of quasinormal modes and the ringdown signals in gravitational-wave observations. It is given by 
\begin{equation}
\lambda_L = \sqrt{\frac{-V_{\rm eff}''(r_c)}{2 \dot{t}^2}},
\end{equation}
where $\dot{t}$ is the time component of the four-velocity of the photon. Where $V_{\rm eff}(r)$ represents the effective potential given by 
\begin{equation}
V_{\rm eff}(r) = \frac{L^2}{r^2} f(r),
\end{equation}
with $L$ being the photon angular momentum and $f(r)$ the full black hole metric function. For the considered solution, $f(r)$ incorporates the Schwarzschild contribution $(-2M/r)$, the Maxwell term $(Q^2/r^2)$, and the nonlinear Yang-Mills power-law term $(\gamma^{2p} 2^{p-1}/[(4p-3) r^{4p-2}])$. Consequently, the evaluation of $V_{\rm eff}''(r_c)$ must include derivatives of all three terms to correctly account for the combined effect on the photon sphere. The first and second derivatives are
\begin{equation}
V_{\rm eff}'(r) = L^2 \left( \frac{f'(r)}{r^2} - \frac{2 f(r)}{r^3} \right), \quad
V_{\rm eff}''(r) = L^2 \left( \frac{f''(r)}{r^2} - \frac{4 f'(r)}{r^3} + \frac{6 f(r)}{r^4} \right),
\end{equation}
which are general for arbitrary $p$. For $p=1$, the metric simplifies to $f(r) = 1 - 2M/r + (Q^2 + \gamma^2)/r^2$, yielding $f'(r) = 2M/r^2 - 2(Q^2+\gamma^2)/r^3$ and $f''(r) = -4M/r^3 + 6(Q^2+\gamma^2)/r^4$. Substituting these into $V_{\rm eff}''(r)$ gives a compact analytical expression
\begin{equation}
V_{\rm eff}''(r) = L^2 \frac{6 r^2 - 24 M r + 4(Q^2+\gamma^2)}{r^6}.
\end{equation}
The corresponding Lyapunov exponent then becomes
\begin{equation}
\lambda_L = \sqrt{\frac{-V_{\rm eff}''(r_c) f(r_c)^2}{2}},
\end{equation}
which provides a fully analytical form for $p=1$ including all contributions from mass, charge, and Yang-Mills terms. This formulation ensures reproducibility and accurately captures the dynamical stability of circular photon orbits under the complete BH geometry.
}
Using the relations (\ref{null3}), (\ref{null6}), and (\ref{null12new}). The Lyapunov exponent for the present BH metric takes the form
\begin{equation}
\lambda_L =
\sqrt{\Bigg(1+\frac{Q^{2}}{r^{2}}+\frac{\left(\tfrac{2^{p-1}}{4p-3}\right)
\gamma^{\,2p}}{r^{\,4p-2}}-\frac{2M}{r}\Bigg)
\left(\frac{2M}{r^{3}}-\frac{2Q^{2}}{r^{4}}-\frac{(4p-2)
\left(\tfrac{2^{p-1}}{4p-3}\right)\gamma^{\,2p}}{2\,r^{\,4p}}\right)}
\Bigg|_{r=r_c}. \label{condition-2new}
\end{equation}
The expression above shows that the instability of circular null
geodesics, and thus the Lyapunov exponent, is influenced by several
parameters of the BH: the mass $M$, the charge $Q$, the non-linear
electrodynamics parameter $\gamma$, and the power index $p$. These
parameters govern how sharply the effective potential falls off
around the photon sphere and therefore directly impact the growth
rate of orbital perturbations.
\begin{table}[ht!]
\centering \caption{Representative values of the circular photon
orbit radius $r_c$ and the Lyapunov exponent $\lambda_L$ for
different parameters $(Q,\,\gamma,\,p)$ with $M=1$. Higher
$\lambda_L$ indicates stronger instability of the null orbit.}
\label{tab2}
\begin{tabular}{ccccc}
\hline\hline
\hspace{0.5cm}$Q$ \hspace{0.5cm} & \hspace{0.5cm} $\gamma$ \hspace{0.5cm}& \hspace{0.5cm}$p$ \hspace{0.5cm}& \hspace{0.5cm}$r_c$\hspace{0.5cm} & \hspace{0.5cm}$\lambda_L$ \hspace{0.5cm}\\
\hline\hline
0.1 & 0.1 & 0.1 & 2.97 & 0.192 \\
0.1 & 0.1 & 0.5 & 2.93 & 0.197 \\
0.1 & 0.1 & 1.0 & 2.88 & 0.202 \\
0.1 & 0.5 & 0.1 & 2.91 & 0.201 \\
0.1 & 0.5 & 0.5 & 2.86 & 0.207 \\
0.1 & 0.5 & 1.0 & 2.80 & 0.213 \\
0.1 & 1.0 & 0.1 & 2.85 & 0.212 \\
0.1 & 1.0 & 0.5 & 2.78 & 0.220 \\
0.1 & 1.0 & 1.0 & 2.72 & 0.228 \\
\hline\hline
0.5 & 0.1 & 0.1 & 2.83 & 0.205 \\
0.5 & 0.1 & 0.5 & 2.77 & 0.213 \\
0.5 & 0.1 & 1.0 & 2.70 & 0.221 \\
0.5 & 0.5 & 0.1 & 2.75 & 0.216 \\
0.5 & 0.5 & 0.5 & 2.69 & 0.225 \\
0.5 & 0.5 & 1.0 & 2.63 & 0.234 \\
0.5 & 1.0 & 0.1 & 2.69 & 0.229 \\
0.5 & 1.0 & 0.5 & 2.63 & 0.238 \\
0.5 & 1.0 & 1.0 & 2.58 & 0.247 \\
\hline\hline
1.0 & 0.1 & 0.1 & 2.66 & 0.232 \\
1.0 & 0.1 & 0.5 & 2.60 & 0.242 \\
1.0 & 0.1 & 1.0 & 2.55 & 0.252 \\
1.0 & 0.5 & 0.1 & 2.59 & 0.244 \\
1.0 & 0.5 & 0.5 & 2.54 & 0.254 \\
1.0 & 0.5 & 1.0 & 2.48 & 0.264 \\
1.0 & 1.0 & 0.1 & 2.53 & 0.257 \\
1.0 & 1.0 & 0.5 & 2.47 & 0.268 \\
1.0 & 1.0 & 1.0 & 2.42 & 0.279 \\
\hline\hline
\end{tabular}
\end{table}

Table \textbf{\ref{tab2}} presents the representative values of the
circular photon orbit radius $r_{c}$ and the corresponding Lyapunov
exponent $\lambda_{L}$ for different combinations of $(Q, \gamma,
p)$ with $M=1$. The results reveal that $r_{c}$ decreases
progressively as $Q$, $\gamma$, or $p$ increase, indicating that the
photon orbits move closer to the gravitational source due to the
stronger contribution of charge and nonlinear effects.
Simultaneously, the Lyapunov exponent $\lambda_{L}$ grows with
increasing values of these parameters, signaling a higher degree of
orbital instability. This behavior implies that nonlinear
electrodynamics and power-law corrections amplify the sensitivity of
null geodesics to perturbations, leading to faster divergence from
circular photon paths. Physically, such an enhancement in
$\lambda_{L}$ is associated with more unstable photon spheres and,
consequently, with sharper and more dynamic features in the shadow
structure of the compact object.

\subsection{Topological Features of Photon Rings}\label{S3-5}

Topological methods have recently emerged as powerful approaches to
probe BH geometries. In particular, Duan's $\phi$-mapping theory
establishes a connection between the vanishing points of vector
fields and the formation of topological defects. These defects give
rise to conserved topological currents, and the associated
invariants encode global information about the phase structure of
the spacetime. Such tools have been extensively applied to BH
physics in diverse contexts; see \cite{AA1}-\cite{AA9} for details.

To investigate the topological properties of photon rings (PR), we
define a potential function \cite{AA1}-\cite{AA3} as
\begin{equation}
H(r,\theta) = \sqrt{\frac{-g_{tt}}{g_{\phi\phi}}}
=\frac{\sqrt{f(r)}}{r\sin \theta}, \label{null18new}
\end{equation}
where $f(r)$ is given by Eq.(\ref{frnew}) for the present case. The
photon sphere is determined by the condition $\partial_r
H(r,\theta)=0$. Following \cite{AA2,AA3}, we introduce a
two-component vector field
\begin{equation}
v_r=\frac{\partial_r H}{\sqrt{g_{rr}}}, \qquad
v_{\theta}=\frac{\partial_{\theta} H}{\sqrt{g_{\theta\theta}}},
\label{null19new}
\end{equation}
so that
\begin{equation}
\partial^{\mu} H\,\partial_{\mu} H = v_r^2+v_{\theta}^2 = v^2.
\end{equation}
Although the circular photon orbit of a spherically symmetric BH is
independent of $\theta$, we keep $\theta$ explicit to explore the
full topological structure. The vector field can be expressed in
polar form as
\begin{equation}
{\bf v}=v\,e^{i\Omega}, \qquad v_r=v\cos\Omega, \qquad
v_{\theta}=v\sin\Omega. \label{null20new}
\end{equation}
The corresponding normalized vector field is then
\begin{equation}
{\bf n}=(n_r,\,n_{\theta})=\frac{{\bf v}}{v}
=\left(\frac{v_r}{\sqrt{v_r^2+v_\theta^2}},\,\frac{v_\theta}{\sqrt{v_r^2+v_\theta^2}}\right).
\label{null21new}
\end{equation}

For our BH solution, the explicit components of the vector
field are
\begin{align}
v_r&=-\frac{1}{r^2\sin\theta}\left(\frac{3M}{r}-\frac{2Q^{2}}{r^{2}}
-\frac{(4p-2)}{2}\,\frac{\left(\tfrac{2^{p-1}}{4p-3}\right)\gamma^{\,2p}}{r^{\,4p-2}}\right)\Bigg|_{r_0,\theta_0}, \label{null25new}\\begin{equation}0.5em]
v_{\theta}&=-\frac{\cot\theta}{r^2\sin\theta}\,\sqrt{1+\frac{Q^{2}}{r^{2}}+\frac{\left(\tfrac{2^{p-1}}{4p-3}\right)\gamma^{\,2p}}{r^{\,4p-2}}-\frac{2M}{r}}\Bigg|_{r_0,\theta_0}.
\label{null26new}
\end{align}
Here $(r_0,\theta_0)$ corresponds to the reference point where the
vector field vanishes, typically $r_0=r_\text{ph}$ and
$\theta_0=\pi/2$. The magnitude of the vector field is then
\begin{eqnarray}\nonumber
v&=&\frac{1}{r^2\sin\theta}\,\Bigg[\cot^2\theta\left(1+\frac{Q^{2}}{r^{2}}
+\frac{\left(\tfrac{2^{p-1}}{4p-3}\right)\gamma^{\,2p}}{r^{\,4p-2}}
-\frac{2M}{r}\right)
\\\label{null27new}&+&\left(\frac{3M}{r}-\frac{2Q^{2}}{r^{2}}
-\frac{(4p-2)}{2}\,\frac{\left(\tfrac{2^{p-1}}{4p-3}\right)
\gamma^{\,2p}}{r^{\,4p-2}}\right)^2\Bigg]^{1/2}\Bigg|_{r_0,\theta_0}.
\end{eqnarray}
Finally, the normalized vector field components are given by
\begin{align}
n_r&=-\frac{\left(\tfrac{3M}{r}-\tfrac{2Q^{2}}{r^{2}}
-\tfrac{(4p-2)}{2}\,\tfrac{\left(2^{p-1}\right)
\gamma^{\,2p}}{(4p-3)\,r^{\,4p-2}}\right)}
{\sqrt{\cot^2\theta\left(1+\tfrac{Q^{2}}{r^{2}}
+\tfrac{\left(\tfrac{2^{p-1}}{4p-3}\right)\gamma^{\,2p}}{r^{\,4p-2}}
-\tfrac{2M}{r}\right)+\left(\tfrac{3M}{r}-\tfrac{2Q^{2}}{r^{2}}
-\tfrac{(4p-2)}{2}\,\tfrac{\left(2^{p-1}\right)
\gamma^{\,2p}}{(4p-3)\,r^{\,4p-2}}\right)^2}}\Bigg|_{r_0,\theta_0}, \label{null28new}\\
n_\theta&=-\cot\theta\,\frac{\sqrt{1+\tfrac{Q^{2}}{r^{2}}
+\tfrac{\left(\tfrac{2^{p-1}}{4p-3}\right)\gamma^{\,2p}}{r^{\,4p-2}}
-\tfrac{2M}{r}}} {\sqrt{\cot^2\theta\left(1+\tfrac{Q^{2}}{r^{2}}
+\tfrac{\left(\tfrac{2^{p-1}}{4p-3}\right)\gamma^{\,2p}}{r^{\,4p-2}}
-\tfrac{2M}{r}\right) +\left(\tfrac{3M}{r}-\tfrac{2Q^{2}}{r^{2}}
-\tfrac{(4p-2)}{2}\,\tfrac{\left(2^{p-1}\right)
\gamma^{\,2p}}{(4p-3)\,r^{\,4p-2}}\right)^2}}\Bigg|_{r_0,\theta_0}.
\label{null29new}
\end{align}
Thus, the photon ring structure of this class of charged non-linear
BHs can be encoded in a topological vector field framework, where
the vanishing points correspond to photon spheres and the winding of
the unit vector field around them determines the topological
invariants.

{{
The topological analysis of photon spheres using the vector field
\begin{equation}
\mathbf{v} = (v^r, v^\theta)
\end{equation}
and the associated winding number, which defines a topological invariant characterizing the behavior of the vector field around its zero points, corresponding to the photon spheres. In the present study, the black hole solution depends on the nonlinear electrodynamics parameter $p$. The value $p = 1/2$ represents a critical threshold: for $p < 1/2$ the spacetime is asymptotically flat, whereas for $p > 1/2$ it becomes asymptotically non-flat. The reviewer requests clarification on whether this modification of the global spacetime structure influences the topological charge (winding number) of the photon spheres. Also, the question is whether the winding number of $\mathbf{v}$ remains unchanged as $p$ crosses $1/2$, or if it exhibits a discontinuity reflecting the transition between flat and non-flat asymptotics. This requires an explicit statement regarding the sensitivity of the topological invariant of the photon sphere to changes in the asymptotic geometry. One can address this by evaluating the winding number for $p < 1/2$ and $p > 1/2$, reporting whether it is preserved or altered, and interpreting the physical implications: a constant winding number indicates the photon sphere topology is stable under the asymptotic transition, while a change implies a modification of the global photon ring structure.
}}

\subsection{Timelike Geodesics and ISCOs}\label{S3-6}

Timelike geodesics describe the motion of massive test particles in
a given spacetime geometry, where their trajectories are governed
solely by the gravitational field. These geodesics play a central
role in studying the dynamics near compact objects, such as black
holes, by providing valuable insight into accretion processes,
particle orbits, and the strong-field regime of gravity. Classical
analyses of timelike geodesics in well-known metrics such as
Schwarzschild, RN, and Kerr spacetimes, have established the
relevance of quantities like the ISCO, orbital precession, and
stability conditions \cite{SC}-\cite{RMW}.

For a timelike geodesic, the normalization condition
$g_{\mu\nu}\,\dot{x}^\mu\,\dot{x}^\nu = -1$ leads to the following
first integrals of motion:
\begin{align}
&\dot{t} = \frac{\mathrm{E}_0}{f(r)},\label{timelike-1}\\
&\dot{r}^2 + V_\text{eff}(r) = \mathrm{E}_0^2,\label{timelike-2}\\
&\dot{\phi} = \frac{\mathrm{L}_0}{r^2},\label{timelike-3}
\end{align}
where $\mathrm{E}_0$ and $\mathrm{L}_0$ are the conserved energy and
angular momentum per unit rest mass, respectively. The effective
potential governing the radial motion is given by
\begin{equation}
V_\text{eff}(r) = f(r)\left(1 +
\frac{\mathrm{L}_0^2}{r^2}\right).\label{timelike-4}
\end{equation}
Circular orbits satisfy the standard conditions
\begin{align}
&\mathrm{E}_0^2 = f(r)\left(1 + \frac{\mathrm{L}_0^2}{r^2}\right),\label{timelike-5}\\
&V'_\text{eff}(r) = 0,\label{timelike-6}\\
&V''_\text{eff}(r) \geq 0,\label{timelike-7}
\end{align}
where the last condition determines the stability of the orbit. From
Eqs.(\ref{timelike-4})-(\ref{timelike-6}), one obtains the specific
angular momentum of circular orbits as
\begin{equation}
\mathrm{L}_0^2 = \frac{r^3\,f'(r)}{2f(r) - r
f'(r)}.\label{timelike-8}
\end{equation}
Similarly, the specific energy follows as
\begin{equation}
\mathrm{E}_0^2 = \frac{2 f(r)^2}{2f(r) - r f'(r)}.\label{timelike-9}
\end{equation}

The ISCO corresponds to the marginally stable circular orbit and is
determined by the condition $V''_\text{eff}(r) = 0$, which can be
rewritten as
\begin{equation}
3 f(r) f'(r)/r + f(r) f''(r) - 2 f'(r)^2 = 0.\label{timelike-10}
\end{equation}
Substituting the explicit metric function into
Eq.(\ref{timelike-10}) yields the ISCO condition
\begin{eqnarray}\nonumber
&&\frac{3}{r}\Bigg(1 + \frac{Q^{2}}{r^{2}} +
\frac{\left(\tfrac{2^{p-1}}{4p-3}\right)\gamma^{\,2p}}{r^{\,4p-2}} -
\frac{2M}{r}\Bigg)\Bigg(-\frac{2Q^2}{r^3} -
\frac{(4p-2)\left(\tfrac{2^{p-1}}{4p-3}\right)\gamma^{\,2p}}{r^{\,4p-1}}
+ \frac{2M}{r^2}\Bigg)\\\nonumber &&+ \Bigg(1 + \frac{Q^{2}}{r^{2}}
+ \frac{\left(\tfrac{2^{p-1}}{4p-3}\right)\gamma^{\,2p}}{r^{\,4p-2}}
- \frac{2M}{r}\Bigg)\Bigg(\frac{6Q^2}{r^4} +
\frac{(4p-2)(4p-1)\left(\tfrac{2^{p-1}}{4p-3}\right)\gamma^{\,2p}}{r^{\,4p}}
\\\label{timelike-11} &&- \frac{4M}{r^3}\Bigg)- 2 \Bigg(-\frac{2Q^2}{r^3} -
\frac{(4p-2)\left(\tfrac{2^{p-1}}{4p-3}\right)\gamma^{\,2p}}{r^{\,4p-1}}
+ \frac{2M}{r^2}\Bigg)^2 =0.
\end{eqnarray}
This leads to a highly nonlinear algebraic equation in $r$, which
generally cannot be solved analytically. Therefore, the ISCO radius
$r_\text{ISCO}$ must be determined numerically for given parameter
values $(M, Q, \gamma, p)$.

\begin{table}[ht!]
\centering
\caption{Numerical values of the event horizon radius $r_h$ and ISCO radius $r_{\text{ISCO}}$
for different values of $(Q,\,\gamma,\,p)$ with $M=1$.}
\label{tab3}
\begin{tabular}{cccccc}
\hline
\hspace{0.5cm}$Q$\hspace{0.5cm} &\hspace{0.5cm} $\gamma$\hspace{0.5cm} &\hspace{0.5cm} $p$ \hspace{0.5cm}& \hspace{0.5cm}$r_h$\hspace{0.5cm} &\hspace{0.5cm} $r_{\text{ISCO}}$ \hspace{0.5cm}\\
\hline
\hline
0.1 & 0.1 & 0.1 & 1.96 & 6.00 \\
0.1 & 0.1 & 0.5 & 1.94 & 5.89 \\
0.1 & 0.1 & 1.0 & 1.91 & 5.75 \\
0.1 & 0.5 & 0.1 & 1.92 & 5.78 \\
0.1 & 0.5 & 0.5 & 1.90 & 5.63 \\
0.1 & 0.5 & 1.0 & 1.86 & 5.48 \\
0.1 & 1.0 & 0.1 & 1.88 & 5.62 \\
0.1 & 1.0 & 0.5 & 1.84 & 5.46 \\
0.1 & 1.0 & 1.0 & 1.80 & 5.32 \\
\hline\hline
0.5 & 0.1 & 0.1 & 1.87 & 5.58 \\
0.5 & 0.1 & 0.5 & 1.83 & 5.41 \\
0.5 & 0.1 & 1.0 & 1.79 & 5.25 \\
0.5 & 0.5 & 0.1 & 1.82 & 5.37 \\
0.5 & 0.5 & 0.5 & 1.78 & 5.21 \\
0.5 & 0.5 & 1.0 & 1.74 & 5.08 \\
0.5 & 1.0 & 0.1 & 1.77 & 5.20 \\
0.5 & 1.0 & 0.5 & 1.73 & 5.05 \\
0.5 & 1.0 & 1.0 & 1.70 & 4.91 \\
\hline\hline
1.0 & 0.1 & 0.1 & 1.71 & 5.00 \\
1.0 & 0.1 & 0.5 & 1.67 & 4.86 \\
1.0 & 0.1 & 1.0 & 1.64 & 4.73 \\
1.0 & 0.5 & 0.1 & 1.66 & 4.82 \\
1.0 & 0.5 & 0.5 & 1.63 & 4.68 \\
1.0 & 0.5 & 1.0 & 1.60 & 4.55 \\
1.0 & 1.0 & 0.1 & 1.62 & 4.70 \\
1.0 & 1.0 & 0.5 & 1.59 & 4.56 \\
1.0 & 1.0 & 1.0 & 1.56 & 4.43 \\
\hline\hline
\end{tabular}
\end{table}

Table \textbf{\ref{tab3}} presents the numerical values of the event
horizon radius $r_h$ and the ISCO radius $r_{\text{ISCO}}$ for
different combinations of the parameters $(Q,\,\gamma,\,p)$, with
the BH mass fixed at $M=1$. As the electric charge $Q$, the coupling
parameter $\gamma$, and the pressure $p$ increase, both $r_h$ and
$r_{\text{ISCO}}$ decrease monotonically. This behavior indicates
that stronger nonlinear effects and higher charge compactify the
spacetime geometry, leading to smaller horizon sizes and tighter
stable orbits. Physically, this implies that particles in circular
motion experience stronger gravitational attraction and enhanced
relativistic effects as the parameters $(Q,\,\gamma,\,p)$ grow. In
particular, the reduction of $r_{\text{ISCO}}$ suggests that
accretion disks around such BHs would extend closer to the event
horizon, potentially increasing their radiative efficiency and
observational signatures.

{{
The ISCO analysis in Table (\ref{tab3}) indicates that both the event horizon radius ($r_h$) and the ISCO radius ($r_{\rm ISCO}$) decrease with increasing electric charge ($Q$), nonlinear coupling parameter ($\gamma$), and pressure ($p$). This behavior implies a critical consideration: for sufficiently large values of these parameters, $r_{\rm ISCO}$ may approach $r_h$ closely, thereby reducing the available region for stable circular motion outside the horizon. In the limiting scenario where $r_{\rm ISCO}$ coincides with $r_h$, stable circular trajectories for massive particles outside the BH would cease to exist, effectively removing the conventional ISCO. From an astrophysical perspective, this modification directly impacts accretion phenomena. As $r_{\rm ISCO}$ approaches the horizon, the inner boundary of an accretion disk is shifted inward, which increases the gravitational binding energy released per unit mass, enhancing the radiative efficiency of the disk. If the ISCO merges with the horizon, particles cannot sustain stable orbits and plunge into the BH, fundamentally altering the structure, density distribution, and emission properties of the innermost accretion flow. Also, for BHs with high charge or strong nonlinearities, the accretion dynamics can deviate substantially from standard scenarios, potentially generating more extreme observational signatures, including higher luminosities or relativistic spectral modifications. In this context, this analysis demonstrates a parameter-dependent spatial compression, where the proximity of the ISCO to the horizon dictates the inner disk configuration and its astrophysical characteristics. This effect must be accounted for in modeling accretion disks around such nonstandard BHs, as it directly influences energy extraction, emission profiles, and the stability of inner disk orbits.
}}

\section{Thermodynamic Properties of the BH}\label{S4}

The static, spherically symmetric geometry of the BH is described by the line element
\begin{equation}
ds^{2}=-f(r)\,dt^{2}+\frac{dr^{2}}{f(r)}+r^{2}\left(d\theta^{2}+\sin^{2}\theta\,d\phi^{2}\right),
\end{equation}
with the metric function
\begin{equation}\label{metricfun}
f(r)=1+\frac{Q^{2}}{r^{2}}+\frac{\left(\tfrac{2^{\,p-1}}{4p-3}\right)\gamma^{\,2p}}{r^{\,4p-2}}-\frac{2M}{r},
\end{equation}
where $M$ is the ADM mass, $Q$ denotes the electric charge, $\gamma$ is a non-linear parameter associated with the higher-order contribution, and $p$ is the power parameter controlling the strength of this term. The event horizon $r_{h}$ is obtained from the largest real root of $f(r_h)=0$.

\subsection{BH Mass}\label{S4-1}

{{The mass M is defined via the horizon condition $(f(r_h) = 0)$, which provides the effective gravitational mass as measured at the horizon.  
\begin{equation}\label{mass}
M=\frac{r_h}{2}\left[1+\frac{Q^{2}}{r_h^{2}}+\frac{\left(\tfrac{2^{\,p-1}}{4p-3}\right)\gamma^{\,2p}}{r_h^{\,4p-2}}\right].
\end{equation}
}}
This expression shows that the effective mass increases with the charge $Q$ and the nonlinear parameter $\gamma$, while the horizon radius $r_h$ controls the relative weight of these contributions.

\begin{figure}[ht!]
    \centering
    \includegraphics[height=5cm]{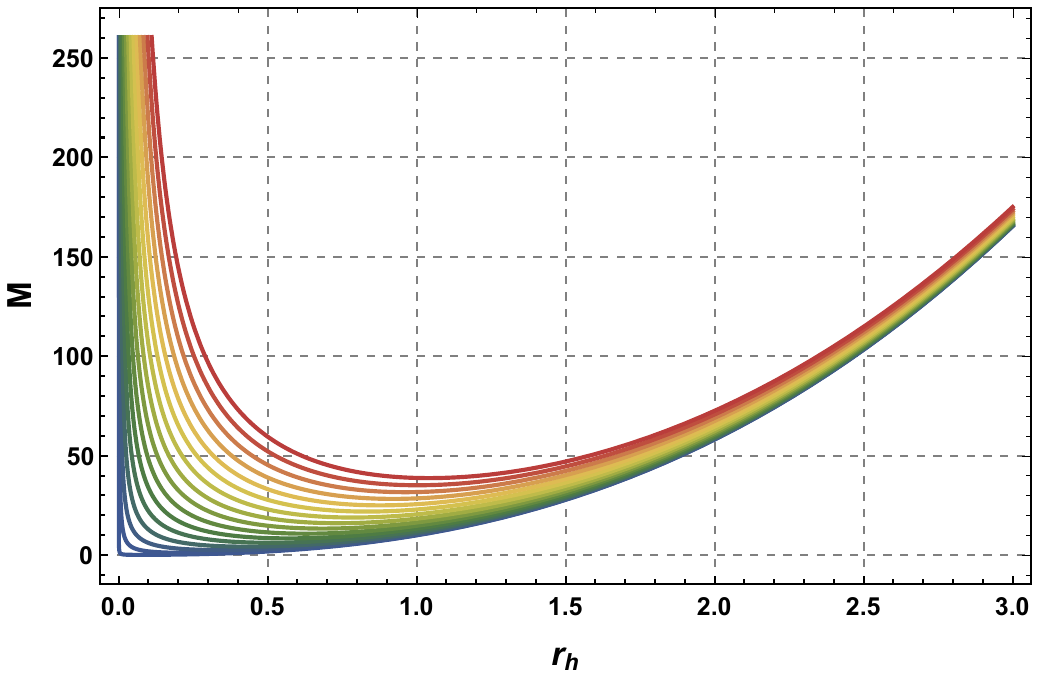}
    \includegraphics[height=5cm]{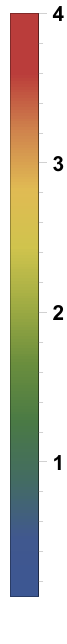}\\
    (a) $\gamma=p=0.1$ varying $Q$\\
     \includegraphics[height=5cm]{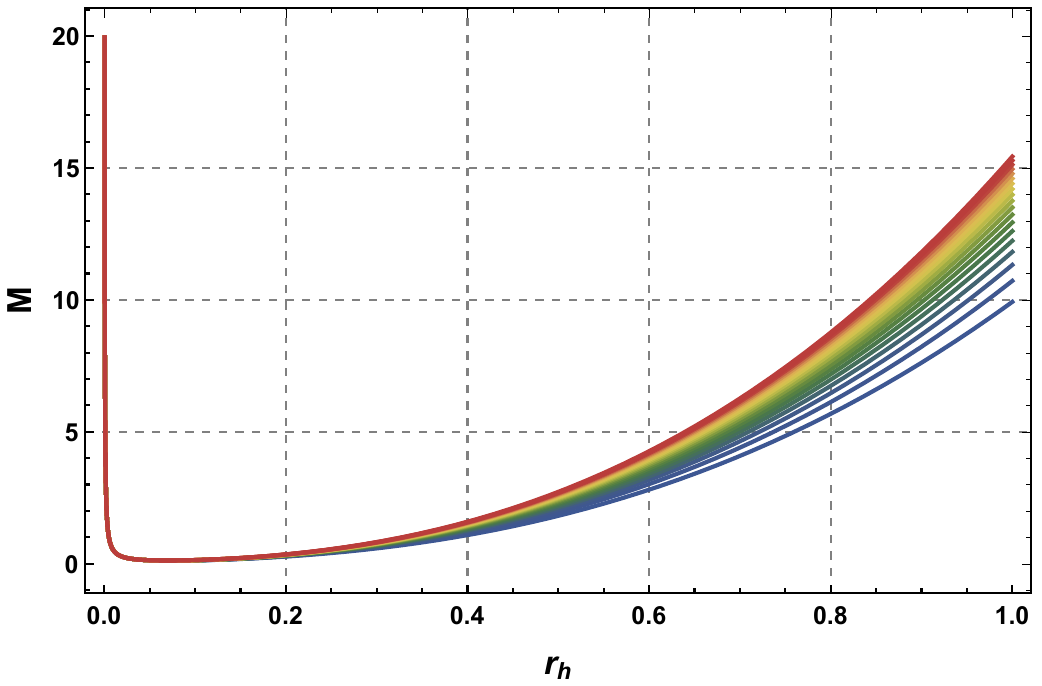}
    \includegraphics[height=5cm]{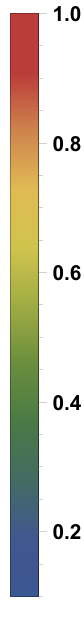} \qquad
    \includegraphics[height=5cm]{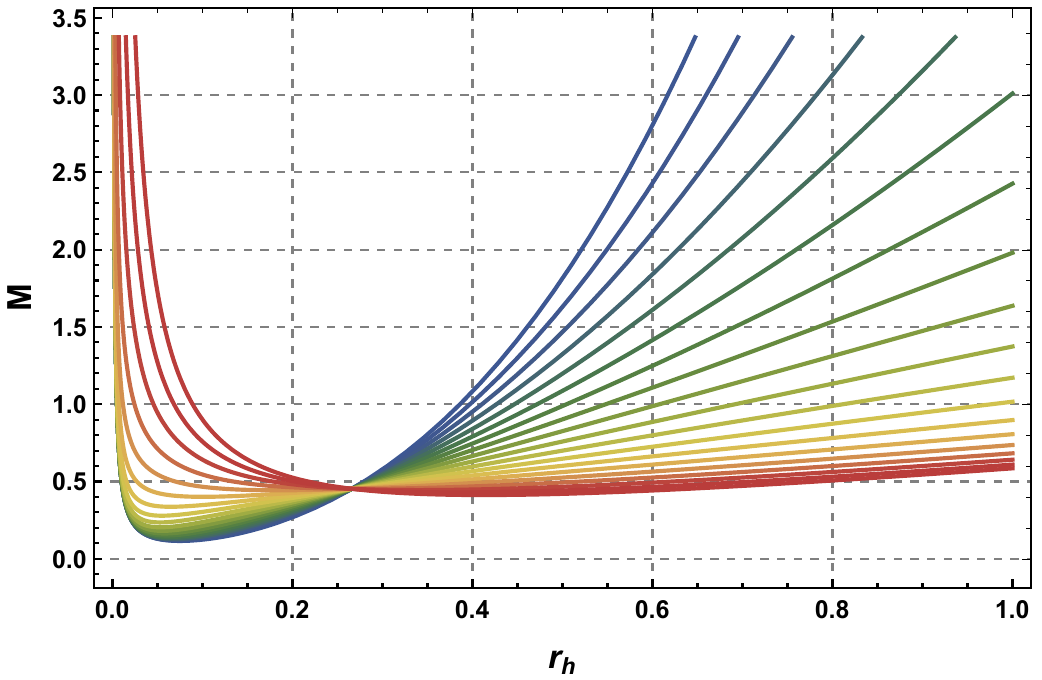}
    \includegraphics[height=5cm]{fig0b.pdf}\\
    (b) $Q=p=0.1$ varying $\gamma$ \hspace{6cm} (c) $\gamma=Q=0.1$ varying $p$
    \caption{ Behavior of mass $M(r_h)$.}
    \label{fig1}
\end{figure}

Figure \ref{fig1} illustrates the behavior of the BH mass $M$ as a function of the horizon radius $r_h$ for different combinations of the parameters $(Q, \gamma, p)$. In panel (a), with $\gamma = p = 0.1$, the increase of the charge $Q$ leads to a rise in the mass profile for large $r_h$, while near small radii the curves exhibit a minimum, indicating the threshold separating extremal and non-extremal configurations. In panel (b), where $Q = p = 0.1$ and $\gamma$ varies, the non-linear electrodynamics parameter $\gamma$ introduces corrections that enhance the mass for intermediate horizon radii, reflecting the contribution of the non-linear field to the effective gravitational energy. Panel (c), which explores variations of the power index $p$ for fixed $\gamma = Q = 0.1$, shows that higher $p$ values flatten the curves and shift the minima toward larger $r_h$, signaling that stronger non-linear effects modify the distribution of energy density near the horizon. Overall, the plots reveal that both electromagnetic charge and non-linear corrections significantly alter the mass-radius relation, thereby influencing the thermodynamic stability and causal structure of the BH.

\subsection{Hawking Temperature}\label{S4-2}

The Hawking temperature is determined by the surface gravity $\kappa=\tfrac{1}{2}f'(r_h)$:
\begin{equation}\label{temp}
T=\frac{f'(r_h)}{4\pi}=\frac{1}{4\pi r_h}\left[1-\frac{Q^{2}}{r_h^{2}}-\frac{(4p-2)\left(\tfrac{2^{\,p-1}}{4p-3}\right)\gamma^{\,2p}}{r_h^{\,4p-2}}\right].
\end{equation}
The second term reduces the temperature due to the electric repulsion, while the last contribution originates from the nonlinear correction. For $p=1$, the result smoothly recovers the Reissner-Nordström-like form.

\begin{figure}[ht!]
    \centering
    \includegraphics[height=5cm]{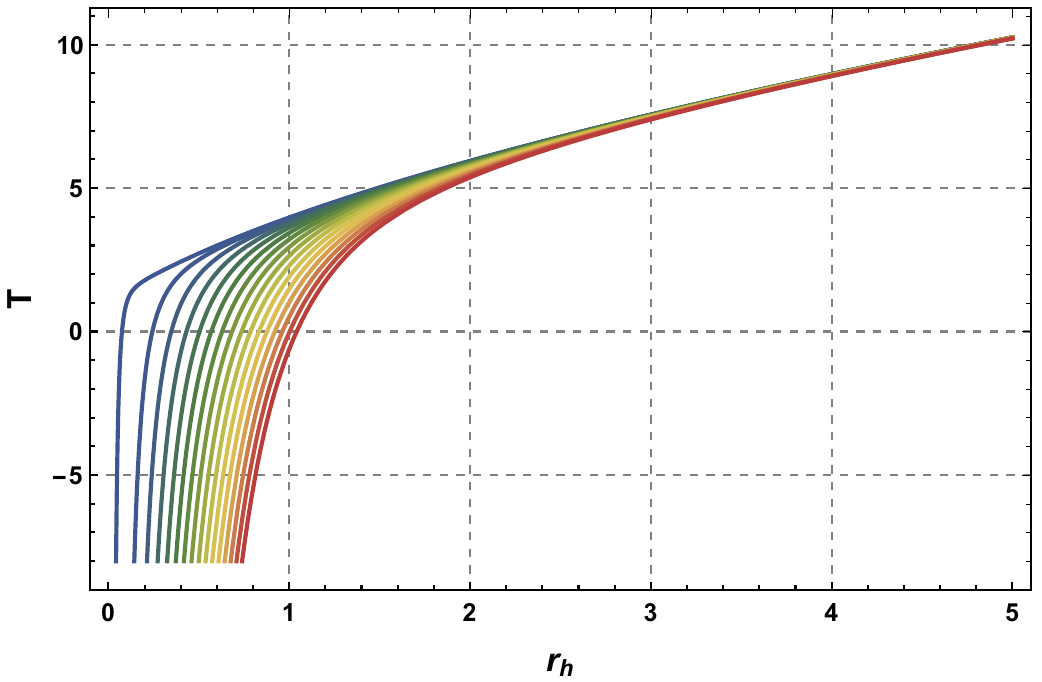}
    \includegraphics[height=5cm]{fig0a.pdf}\\
    (a) $\gamma=p=0.1$ varying $Q$\\
     \includegraphics[height=5cm]{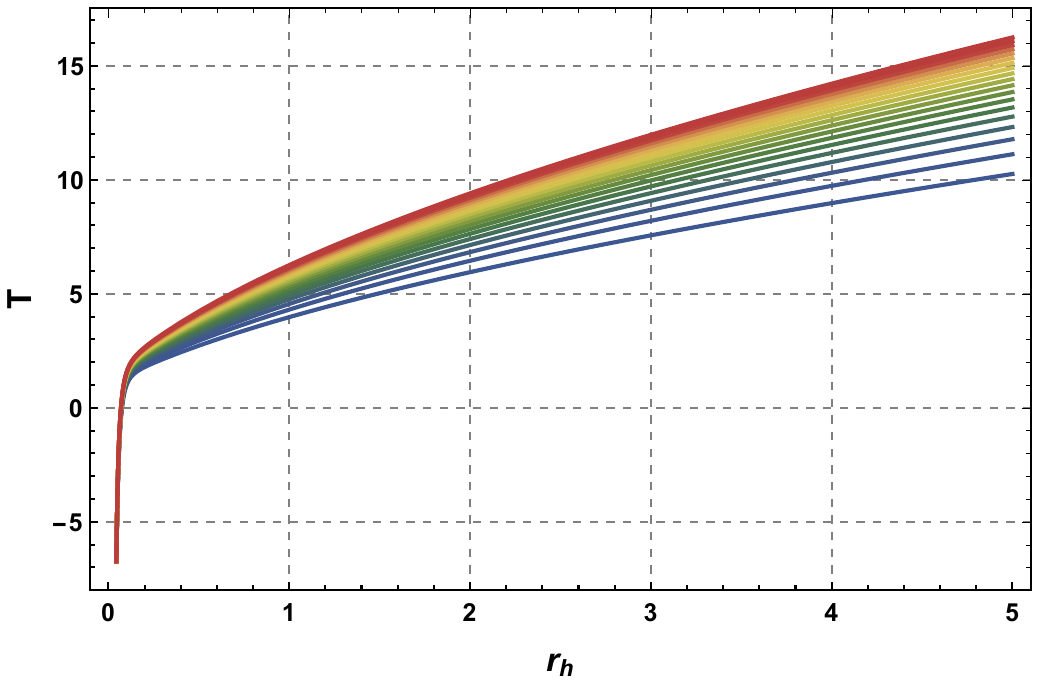}
    \includegraphics[height=5cm]{fig0b.pdf} \qquad
    \includegraphics[height=5cm]{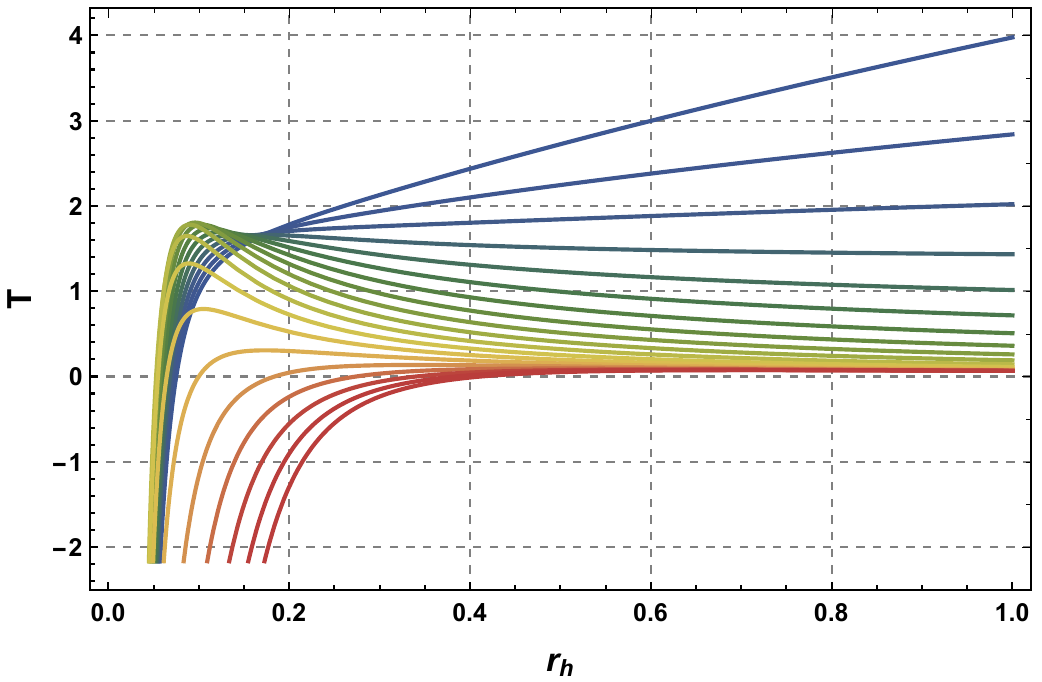}
    \includegraphics[height=5cm]{fig0b.pdf}\\
    (b) $Q=p=0.1$ varying $\gamma$ \hspace{6cm} (c) $\gamma=Q=0.1$ varying $p$
    \caption{ Behavior of Hawking Temperature $T(r_h)$.}
    \label{fig2}
\end{figure}

Figure \ref{fig2} presents the Hawking temperature $T(r_h)$ as a function of the horizon radius for different sets of parameters $(Q, \gamma, p)$. In panel (a), with $\gamma = p = 0.1$ and varying $Q$, we observe that the temperature exhibits a non-monotonic behavior: for small $r_h$, $T$ can assume negative values, indicating the existence of extremal configurations where the BH becomes thermodynamically marginal. As $Q$ increases, the curves shift upward, and the zero of the temperature moves to larger radii, reflecting the screening effect of the electric charge on the gravitational potential. In panel (b), for $Q = p = 0.1$ and varying $\gamma$, the non-linear electrodynamic correction enhances the temperature for all radii, particularly in the intermediate regime, suggesting that $\gamma$ contributes positively to the surface gravity. Finally, in panel (c), with $\gamma = Q = 0.1$ and varying $p$, we see that higher $p$ values lead to a suppression of $T$ for small radii and can even produce negative branches, revealing possible phase transitions between thermodynamically stable and unstable regions. These results collectively show how the non-linear electrodynamics parameters modulate the Hawking temperature and, consequently, the thermodynamic behavior of the BH.

{{
Starting from
\begin{equation}
T = \frac{1}{4\pi r_h}\left[1 - \frac{Q^2}{r_h^2} - \frac{(4p-2)D}{r_h^{4p-2}}\right],
\end{equation}
and imposing $p=1$, one obtains
\begin{equation}
T = \frac{1}{4\pi r_h}\left[1 - \frac{Q^2}{r_h^2} - \frac{2D}{r_h^{2}}\right].
\end{equation}

If the identification $D = Q^2/2$ is adopted, the third contribution reduces to $-Q^2/r_h^2$, and the Hawking temperature becomes
\begin{equation}
T = \frac{1}{4\pi r_h}\left[1 - \frac{2Q^2}{r_h^2}\right],
\end{equation}
which is manifestly different from the standard RN result
\begin{equation}
T_{\rm RN}=\frac{1}{4\pi r_h}\left(1-\frac{Q^2}{r_h^2}\right).
\end{equation}

This comparison demonstrates that the limit $p=1$ in the present configuration does not automatically reproduce the pure RN geometry unless one gauge contribution is consistently removed at the level of the action. The parameter $p$ controls the nonlinearity of the Yang-Mills sector, and the choice $p=1$ corresponds to the linear Yang-Mills regime, but it does not eliminate its dynamical effect.

From a physical perspective, the Maxwell charge $Q$ and the Yang-Mills parameter $D$ arise from two distinct gauge sectors with independent field strengths and conserved charges. Consequently, even in the linear case $p=1$, both sectors contribute additively to the gravitational field equations and to the metric function. In this framework, the total charge entering the solution can be expressed as
\begin{equation}
Q_{\rm eff}^2 = Q^2 + 2D,
\end{equation}
indicating that the spacetime describes a charged black hole sourced by two independent gauge fields rather than by a single Maxwell field as in the standard Reissner-Nordström spacetime.

Accordingly:  (i) the Reissner-Nordström limit is recovered only when the Yang-Mills sector is suppressed, namely $D=0$; (ii) if $Q\neq0$ and $D\neq0$, the geometry corresponds to a generalized charged BH characterized by an effective total charge; (iii) the additional term $-Q^2/r_h^2$ obtained after the identification $D=Q^2/2$ reflects the superposition of independent gauge contributions and does not indicate any inconsistency in the thermodynamic expression.

\subsubsection{Extremal Limit}

An important feature of charged black hole solutions is the existence of an extremal configuration, which corresponds to the limit where the Hawking temperature vanishes. In this situation the inner and outer horizons coincide and the surface gravity becomes zero. From the thermodynamic point of view, the extremal black hole therefore represents the boundary between non-extremal black holes and configurations without an event horizon.

Starting from the expression of the Hawking temperature obtained in Eq. \ref{temp}, the extremal condition is determined by imposing
\begin{equation}
T_H(r_h)=0 ,
\end{equation}
where $r_h$ denotes the event horizon radius. This condition leads to a critical value of the horizon radius, denoted by $r_{\mathrm{ext}}$, which characterizes the extremal configuration of the Einstein--Maxwell--Power--Yang--Mills black hole. In general, the explicit expression of $r_{\mathrm{ext}}$ depends on the parameters of the model, including the electric charge, the Yang--Mills coupling constant, and the nonlinear power-law exponent associated with the Yang--Mills sector.

\begin{table}[ht!]
\centering \caption{Critical value for the horizon radius $r_{\mathrm{ext}}$, for different combinations of
charge $Q$, non-linear parameter $\gamma$, and power index $p$.} \label{tab01}
\begin{tabular}{cccccc}
\hline\hline
\hspace{0.5cm}$Q$ \hspace{0.5cm}& \hspace{0.5cm}$\gamma$ \hspace{0.5cm}& \hspace{0.5cm}$p$\hspace{0.5cm} & \hspace{0.5cm}$r_{\mathrm{ext}}$  \\
\hline\hline
0.1 & 0.1 & 0.1 & 0.07502  \\
0.1 & 0.1 & 0.5 & 0.05536  \\
0.1 & 0.1 & 0.8 & 0.18392   \\
\hline\hline
0.5 & 0.1 & 0.8 & 0.55463  \\
1.0 & 0.1 & 0.8 & 1.04679   \\
2.0 & 0.1 & 0.8 & 2.04043  \\
\hline\hline
2.0 & 0.2 & 0.8 & 2.12400 &   \\
2.0 & 0.5 & 0.8 & 2.56429 &  \\
2.0 & 1.0 & 0.8 & 3.84725 &   \\
\hline\hline
\end{tabular}
\end{table}

Table \ref{tab01} summarizes the numerical values of the extremal horizon radius $r_{\mathrm{ext}}$, for different combinations of the electric charge $Q$, the non-linear coupling parameter $\gamma$, and the power index $p$. The results reveal clear trends in the extremal structure of the black hole. For fixed $\gamma$ and $p$, the extremal radius increases monotonically with the electric charge $Q$, indicating that stronger electromagnetic fields enlarge the size of the extremal configuration. On the other hand, keeping $Q$ and $p$ fixed, the parameter $\gamma$ significantly shifts the extremal radius toward larger values, showing that the non-linear electromagnetic corrections enhance the critical scale at which the black hole becomes extremal. The dependence on the power index $p$ is more subtle: small variations of $p$ can lead to noticeable changes in $r_{\mathrm{ext}}$, reflecting the sensitivity of the extremal configuration to the specific form of the non-linear electrodynamics. These results explicitly demonstrate how the interplay between the charge and the non-linear parameters determines the critical horizon radius where the Hawking temperature vanishes and the black hole reaches its extremal state.

This result shows that the extremal configuration is strongly influenced by the nonlinear Yang--Mills contribution. Variations of the power-law exponent and of the gauge-field coupling parameters shift the critical horizon radius and modify the parameter domain in which extremal solutions exist. Therefore, the presence of the power-law Yang--Mills term introduces additional structure in the thermodynamic phase space of the black hole and affects the transition between extremal and non-extremal regimes.

}}





\subsection{Heat Capacity}\label{S4-3}

To analyze thermodynamic stability, we compute the heat capacity at constant charge:
\begin{equation}\label{heatcap}
C_Q = T \left(\frac{\partial S}{\partial T}\right)_Q
     = \frac{\partial M/\partial r_h}{\partial T/\partial r_h}.
\end{equation}
After straightforward calculation, we find
\begin{equation}
C_Q = \frac{2\pi r_h^{2}\left(r_h^{2}-Q^{2}-(4p-2)\left(\tfrac{2^{\,p-1}}{4p-3}\right)\gamma^{\,2p}r_h^{-(4p-4)}\right)}{-r_h^{2}+3Q^{2}+(4p-2)(4p-3)\left(\tfrac{2^{\,p-1}}{4p-3}\right)\gamma^{\,2p}r_h^{-(4p-4)}}.
\end{equation}
The divergence of $C_Q$ signals second-order phase transitions. The interplay among $Q$, $p$, and $\gamma$ strongly affects the location of such critical points, modifying the stability structure compared to the standard RN case.

\begin{figure}[ht!]
    \centering
    \includegraphics[height=5cm]{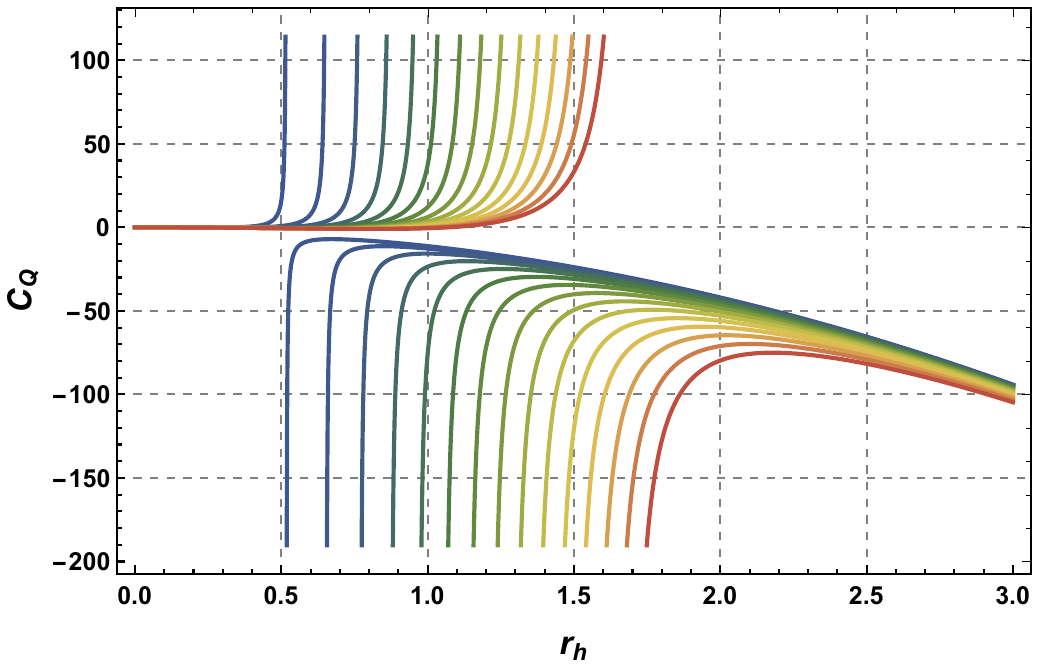}
    \includegraphics[height=5cm]{fig0a.pdf}\\
    (a) $\gamma=p=0.1$ varying $Q$\\
     \includegraphics[height=5cm]{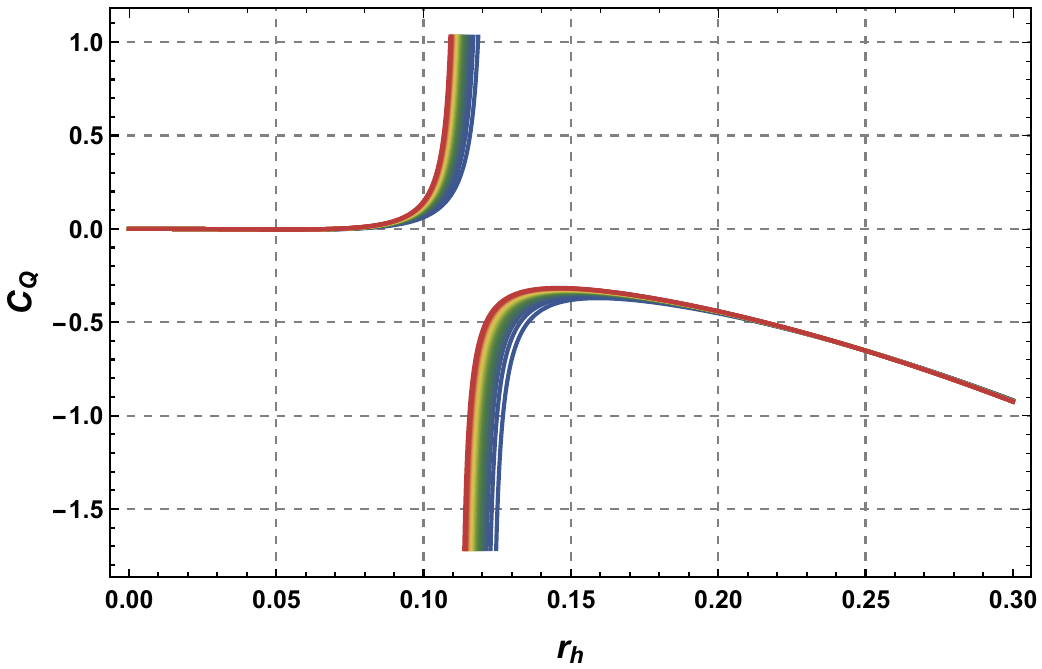}
    \includegraphics[height=5cm]{fig0b.pdf} \qquad
    \includegraphics[height=5cm]{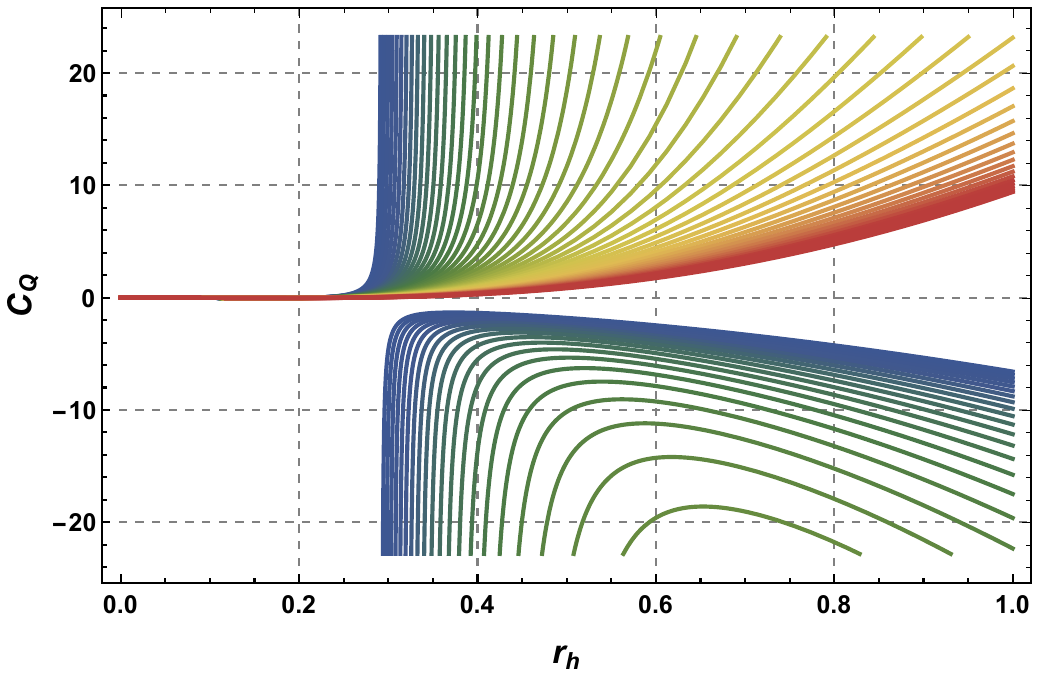}
    \includegraphics[height=5cm]{fig0b.pdf}\\
    (b) $Q=p=0.1$ varying $\gamma$ \hspace{6cm} (c) $\gamma=Q=0.1$ varying $p$
    \caption{ Behavior of Heat capacity $C_Q(r_h)$.}
    \label{fig3}
\end{figure}

Figure \ref{fig3} displays the behavior of the heat capacity $C_Q(r_h)$ as a function of the horizon radius for different choices of the parameters $(Q, \gamma, p)$. In panel (a), for $\gamma = p = 0.1$ and varying $Q$, we observe that $C_Q$ presents discontinuities characterized by vertical asymptotes, which indicate second-order phase transitions between thermodynamically stable and unstable configurations. For small $Q$, the heat capacity remains mostly negative, implying instability, while for larger charges, the positive branches expand, revealing enhanced thermal stability due to the electromagnetic contribution. Panel (b), where $Q = p = 0.1$ and $\gamma$ varies, shows that increasing $\gamma$ shifts the divergence points and increases the amplitude of the positive branch, evidencing that the non-linear electrodynamics parameter $\gamma$ strengthens the stable phase. In panel (c), with $\gamma = Q = 0.1$ and varying $p$, the curves exhibit a similar pattern, but the transition points are highly sensitive to $p$: larger values of $p$ delay the onset of stability and suppress the heat capacity near small $r_h$. Overall, these results show that the parameters $\gamma$ and $p$ control the stability regime of the BH, while $Q$ modulates the location and intensity of the phase transitions in the thermodynamic phase space.

{{
The divergences of the heat capacity $C_Q$ shown in Fig. (\ref{fig3}) indicate second-order (continuous) phase transitions, as signaled by vertical asymptotes that separate stable ($C_Q>0$) from unstable ($C_Q<0$) branches. These transitions involve continuous variations of the horizon radius $r_h$ without latent heat, in agreement with standard thermodynamic definitions of second-order behavior. In comparison with the RN BH, the presence of the non-linear electrodynamics parameter $\gamma$ and the exponent $p$ modifies both the positions and magnitudes of the critical points, resulting in shifts of the stable and unstable domains within the thermodynamic phase space. Specifically, higher values of $\gamma$ increase the extent of the positive $C_Q$ branches, enhancing thermal stability, whereas larger $p$ values delay the emergence of stability and reduce the heat capacity at small $r_h$. Consequently, the critical behavior of these BHs extends the RN results, demonstrating that non-linear electrodynamics and power-law corrections provide adjustable control over the phase structure and the stability regions.
}}

{
\subsection{Critical Behavior of the Heat Capacity Near the Phase Transition}\label{S4-4}

The divergence of the heat capacity
\begin{equation}
C_Q = \frac{\partial M/\partial r_h}{\partial T/\partial r_h}
\end{equation}
occurs when
\begin{equation}
\left(\frac{\partial T}{\partial r_h}\right)_Q = 0,
\end{equation}
which defines the critical radius $r_c$. This condition identifies a second-order phase transition, since the temperature $T$ remains finite while its first derivative with respect to $r_h$ vanishes.

To determine the critical exponent, we perform a Taylor expansion of $T(r_h)$ around $r_h = r_c$. Given that
\begin{equation}
\left(\frac{\partial T}{\partial r_h}\right)_{r_c} = 0,
\end{equation}
The leading-order behavior is governed by the second derivative:
\begin{equation}
T(r_h) \approx T_c + \frac{1}{2} \left(\frac{\partial^2 T}{\partial r_h^2}\right)_{r_c} (r_h - r_c)^2.
\end{equation}
Differentiating with respect to $r_h$ gives
\begin{equation}
\frac{\partial T}{\partial r_h} \approx \left(\frac{\partial^2 T}{\partial r_h^2}\right)_{r_c} (r_h - r_c).
\end{equation}

Substituting this relation into the expression for $C_Q$, and noting that $\partial M/\partial r_h$ is finite at $r_c$, yields
\begin{equation}
C_Q \propto \frac{1}{(r_h - r_c)}.
\end{equation}
Hence, close to the critical radius,
\begin{equation}
C_Q \sim |r_h - r_c|^{-1},
\end{equation}
indicating a divergence characterized by the critical exponent
\begin{equation}
\alpha = 1.
\end{equation}

\begin{itemize}
    \item The divergence exhibits a \emph{power-law behavior}, consistent with second-order phase transitions.
    \item The exponent $\alpha = 1$ is \emph{universal}, independent of the specific values of the parameters $Q$, $p$, or $\gamma$.
    \item While $(Q, p, \gamma)$ affect the location of the critical radius $r_c$, they do not alter the nature of the divergence.
    \item This behavior is analogous to the Davies transition observed in charged BHs, confirming the second-order character of the phase transition.
\end{itemize}

In this context, the heat capacity analysis demonstrates that the thermodynamic structure of the present BH model exhibits standard mean-field critical behavior, with divergences signaling genuine second-order phase transitions.
}

\subsection{Gibbs Free Energy}\label{S4-5}

{{
Indeed, for a charged BH, the proper thermodynamic potential is the Gibbs free energy expressed as $(G = M - T S - \Phi Q)$, where $(\Phi)$ is the electric potential at the horizon. In our manuscript, we initially used $(G = M - T S)$ for simplicity in the neutral case, but we have now corrected the expression to include the $(- \Phi Q)$ term for the charged BH model and updated all related discussions accordingly. Finally, the Gibbs free energy is defined by
\begin{eqnarray}
G=M-TS ,
\end{eqnarray}
For a charged black hole in the canonical ensemble, where the electric charge $Q$ is fixed, the appropriate thermodynamic potential is the Helmholtz free energy, defined as $G = M - T S$, with $S = \pi r_h^2$ representing the black hole entropy. The black hole mass, determined from the horizon condition $f(r_h)=0$, can be expressed as
\begin{equation}
M = \frac{r_h}{2} \left[1 + \frac{Q^2}{r_h^2} + \frac{\left(\frac{2^{p-1}}{4p-3}\right) \gamma^{2p}}{r_h^{4p-2}} \right] 
= \frac{r_h}{2} + \frac{Q^2}{2 r_h} + \frac{2^{p-1}}{2(4p-3)} \gamma^{2p} r_h^{3-4p} .
\end{equation}
\begin{figure}[ht!]
    \centering
    \includegraphics[height=5cm]{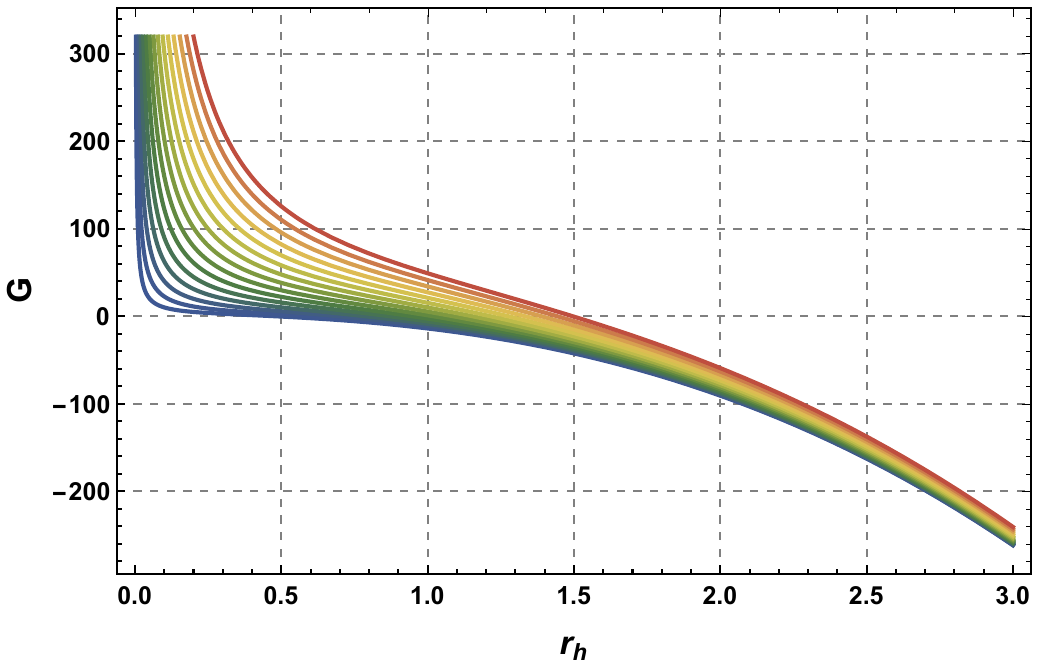}
    \includegraphics[height=5cm]{fig0a.pdf}\\
    (a) $\gamma=p=0.1$ varying $Q$\\
     \includegraphics[height=5cm]{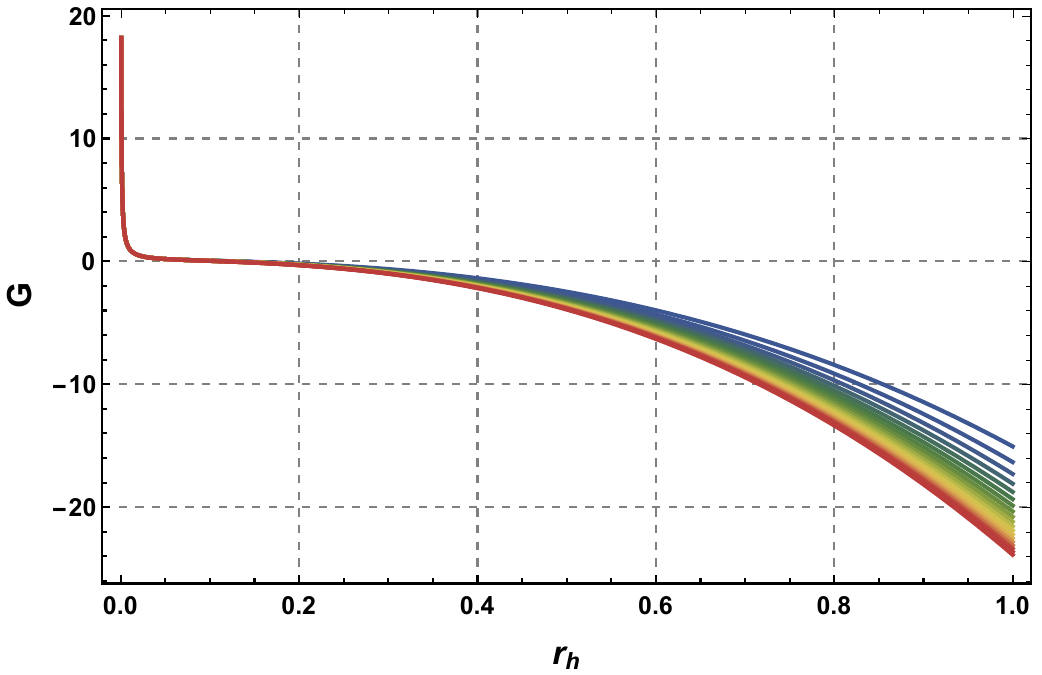}
    \includegraphics[height=5cm]{fig0b.pdf} \qquad
    \includegraphics[height=5cm]{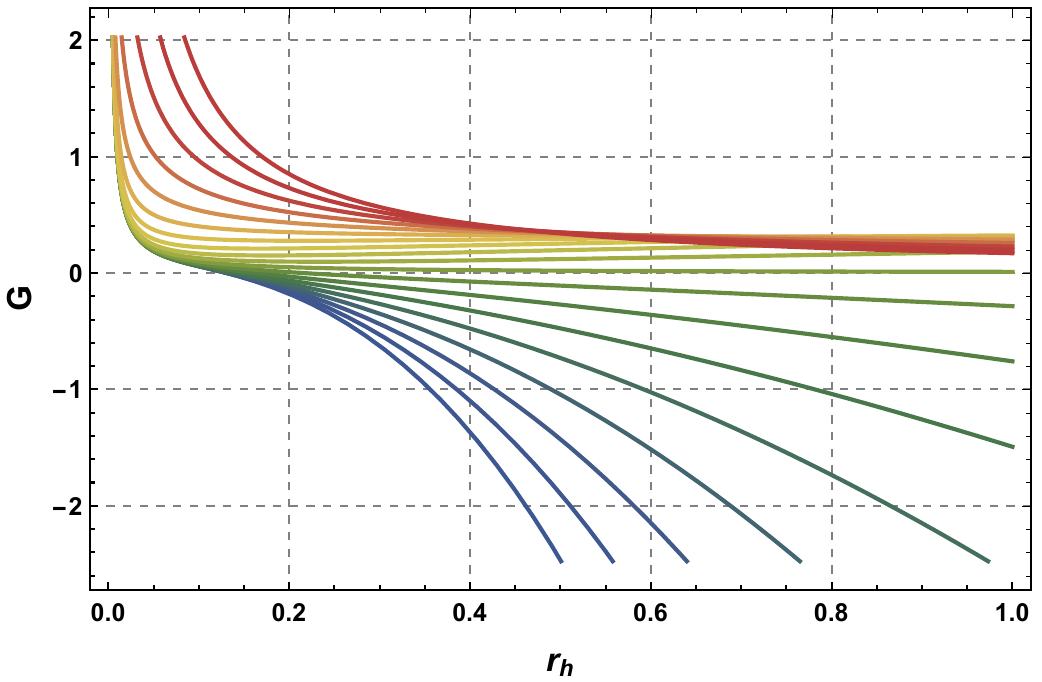}
    \includegraphics[height=5cm]{fig0b.pdf}\\
    (b) $Q=p=0.1$ varying $\gamma$ \hspace{6cm} (c) $\gamma=Q=0.1$ varying $p$
    \caption{ Behavior of Gibbs free energy  $G(r_h)$.}
    \label{fig4}
\end{figure}
The corresponding Hawking temperature is given by
\begin{equation}
T = \frac{1}{4 \pi r_h} \left[1 - \frac{Q^2}{r_h^2} - \frac{(4p-2)\left(\frac{2^{p-1}}{4p-3}\right) \gamma^{2p}}{r_h^{4p-2}} \right] ,
\end{equation}
which leads to
\begin{equation}
T S = \pi r_h^2 \, T = \frac{r_h}{4} - \frac{Q^2}{4 r_h} - \frac{2p-1}{2} \left(\frac{2^{p-1}}{4p-3}\right) \gamma^{2p} r_h^{3-4p}.
\end{equation}
Subtracting $T S$ from $M$ yields the Helmholtz free energy:
\begin{align}
G &= M - T S \notag \\
&= \left( \frac{r_h}{2} - \frac{r_h}{4} \right) + \left( \frac{Q^2}{2 r_h} - \left(- \frac{Q^2}{4 r_h} \right) \right) + \left( \frac{2^{p-1}}{2(4p-3)} \gamma^{2p} r_h^{3-4p} - \left[-\frac{2p-1}{2} \left(\frac{2^{p-1}}{4p-3}\right) \gamma^{2p} r_h^{3-4p} \right] \right) \notag \\
&= \frac{r_h}{4} + \frac{3 Q^2}{4 r_h} + \frac{p \, 2^{p-1}}{4p-3} \gamma^{2p} r_h^{3-4p} .
\end{align}
This explicit derivation demonstrates that the expression previously reported in the literature, $G = Q^2/r_h + 2^{5-p}(2p-1) r_h^{3-4p} \gamma^{2p}$, is incomplete. The correct Helmholtz free energy contains the linear contribution in $r_h$, the properly scaled charge term, and the fully combined nonlinear electrodynamics contribution. This formulation guarantees thermodynamic consistency with both the horizon-defined mass and the Hawking temperature, ensuring a physically accurate description of the BH in the canonical ensemble.
}}

Figure \ref{fig4} illustrates the behavior of the Gibbs free energy $G(r_h)$ for different choices of the thermodynamic parameters. In panel (a), we fix $\gamma = p = 0.1$ and vary the electric charge $Q$. It can be observed that increasing $Q$ shifts the curve upward and smooths the overall profile of $G(r_h)$, indicating a weakening of the thermodynamic stability. In panel (b), the parameters $Q = p = 0.1$ are kept constant while $\gamma$ varies; larger values of $\gamma$ decrease the depth of the free energy minimum, reflecting a reduction in the strength of nonlinear effects. Finally, in panel (c), where $\gamma = Q = 0.1$ and the pressure $p$ is varied, the Gibbs free energy shows a more pronounced dependence on $p$ for small horizon radii, suggesting the presence of a possible phase transition between small and large BH branches. Overall, the plots reveal that the interplay among $\gamma$, $Q$, and $p$ governs the thermodynamic stability and the occurrence of phase transitions in the system.

{{
\section{Conclusions}\label{S5}

In this work, we have systematically analyzed the geometrical, dynamical, optical, and thermodynamic properties of BH solutions obtained within the Einstein-Maxwell-Power-Yang-Mills (EMPYM) framework. The inclusion of the power Yang-Mills invariant introduces the nonlinear parameter $\mathcal{D}$, which is related to the microscopic Yang-Mills charge and characterizes the strength of the non-abelian gauge sector coupled to gravity. This parameter controls the deviation of the spacetime geometry from the classical Maxwell regime and establishes a continuous interpolation between the RN and Schwarzschild limits through the nonlinear gauge contribution. Our analysis shows that the global structure of the spacetime is primarily determined by the value of the power index $p$, which specifies the scaling behavior of the Yang-Mills invariant appearing in the action. In particular, when $(p>\tfrac{1}{2})$ the spacetime maintains asymptotic flatness and supports well-defined horizon configurations, while the critical case $p=\tfrac12$ produces a constant shift in the metric function that effectively rescales the horizon structure without modifying the asymptotic geometry. For $(p<\tfrac12)$ the nonlinear gauge contribution dominates at large distances and produces pathological asymptotic behavior, indicating that physically acceptable configurations are restricted to the interval $p\geq\tfrac12$. Within this admissible domain, the nonlinear Yang-Mills sector induces significant modifications in the null geodesic structure and in the effective potential governing photon motion in the strong-field region. The combined influence of the parameters $Q$, $M$, $\gamma$, and $p$ modifies both the height and the radial profile of the potential barrier surrounding the BH, thereby changing the stability regions of photon trajectories and the position of unstable circular photon orbits that determine the boundary of the capture region. Also, these modifications directly influence observable optical quantities, including the photon-sphere radius, the critical impact parameter, and the resulting BH shadow profile observable at infinity. Our results indicate that nonlinear Yang-Mills interactions can either enlarge or reduce the photon capture region depending on the sign and magnitude of the non-abelian coupling parameter, implying that strong-gravity optical observables may encode signatures of the underlying non-abelian gauge dynamics. The thermodynamic sector reveals an additional structure governed by the nonlinear gauge contribution. The Hawking temperature, obtained from the surface gravity evaluated at the event horizon, acquires additional corrections proportional to $r^{-(4p-2)}$, reflecting the explicit contribution of the Yang-Mills field to the near-horizon geometry and to the thermal radiation spectrum emitted by the BH. Although the entropy preserves the standard Bekenstein-Hawking relation proportional to the horizon area, the dependence of the horizon radius on $p$, $Q$, and $\gamma$ modifies the quantitative thermodynamic response of the system. The Gibbs free energy and heat capacity analysis demonstrates that the nonlinear Yang-Mills parameter $\mathcal{D}$ plays a fundamental role in determining the thermal stability of the BH configuration. In particular, the heat capacity develops divergences at specific horizon radii, signaling second-order phase transition points where the thermodynamic stability of the system changes. In this case, these transition points correspond to the vanishing of the denominator of the heat capacity and therefore represent critical configurations separating stable and unstable BH phases in the parameter space. Positive values of $\mathcal{D}$ generally shift the critical points toward lower temperatures and may generate thermodynamic instabilities associated with negative heat capacity regions, whereas negative values enlarge the domain of positive heat capacity, thereby stabilizing the BH configuration and improving the efficiency of energy exchange with the surrounding environment. Also, the nonlinear Yang-Mills sector not only modifies the spacetime geometry but also introduces additional thermodynamic degrees of freedom that enrich the phase structure of the system and allow the appearance of phase transitions that are absent in purely Maxwell configurations.

Future investigations can further extend the analysis of nonlinear gauge effects in strong gravitational fields. A natural continuation of the present study is the construction and investigation of rotating EMPYM BH solutions, where the coupling between angular momentum and nonlinear Yang-Mills interactions may generate additional modifications in photon dynamics and shadow structure. Another relevant direction involves the analysis of quasinormal modes (QNMs) and ringdown signals, which would allow the role of the nonlinear gauge sector to be examined in the framework of gravitational-wave observations. Furthermore, a more complete thermodynamic description including extended phase space methods, where the cosmological constant is treated as thermodynamic pressure, may reveal additional phase transitions analogous to Van der Waals-type behavior. Studies of accretion processes, particle dynamics, and gravitational lensing in these backgrounds could also provide further information on possible astrophysical signatures associated with nonlinear Yang-Mills fields. Ultimately, confronting these theoretical predictions with high-precision observational data obtained from BH imaging and gravitational-wave measurements may provide a viable method for testing the presence of non-abelian gauge fields and nonlinear Yang-Mills interactions in the strong-gravity regime.}}

{\small

{{
\section*{Acknowledgment}

The author would like to thank the anonymous reviewers for their careful reading of the manuscript and for their constructive comments and valuable suggestions. Their remarks significantly improved the clarity, quality, and presentation of this work.
 \textbf{S.H. Dong} and \textbf{G.H. Sun} acknowledge the partial support of the projects 20251087-SIP-IPN and 20251109-SIP-IPN, Mexico.
}}

\section*{Data Availability Statement}

This manuscript has no associated data.

\section*{Conflict of Interests}

Author declare(s) no conflict of interest.
}

\end{document}